\documentclass[preprint,showpacs,preprintnumbers,amsmath,amssymb,floats,groupedaddress]{revtex4-1} 
\usepackage{graphicx}
\usepackage[english]{babel}
\usepackage{amsmath}
\usepackage{amssymb}
\usepackage{color}

\newcommand{\be}{\begin{eqnarray}} 
\newcommand{\ee}{\end{eqnarray}}

\begin{document}
\title{Quantum criticality in the two-dimensional dissipative quantum XY model - II}
\author{Lijun Zhu}
\affiliation{Department of Physics and Astronomy, University of California, Riverside CA 92521, USA}
\author{Changtao Hou}
\affiliation{Department of Physics and Astronomy, University of California, Riverside CA 92521, USA}
\author{Chandra M. Varma}
\affiliation{Department of Physics and Astronomy, University of California, Riverside CA 92521, USA}

\begin{abstract}
Earlier Monte-Carlo calculations on the dissipative two-dimensional XY model are extended in several directions. We study the phase diagram and the correlation functions when dissipation is very small, where it has properties of the classical 3D-XY transition, i.e. one with a dynamical critical exponent $z=1$.  The transition changes from $z=1$ to the class of criticality with $z \to \infty$ driven by topological defects, discovered earlier, beyond a critical dissipation. We also find that the critical correlations have power-law singularities as a function of tuning the ratio of the kinetic energy to the potential energy for fixed large dissipation, as opposed to essential singularities on tuning dissipation keeping the former fixed. A phase with temporal disorder but spatial order of the Kosterlitz-Thouless form is also further investigated. We also present results for the transition when the allowed Caldeira-Leggett form of dissipation and the allowed form of dissipation coupling to the compact rotor variables are both included. The nature of the transition is then determined by the Caldeira-Leggett form.
\end{abstract}
\maketitle

\section{Introduction}
\label{sec:intro}

In a previous paper \cite{ZhuChenCMV2015} (referred to as I), we have investigated the dissipative quantum XY (DQXY) model  by quantum Monte-Carlo calculations in two spatial dimensions (2D) for a range of parameters. This followed an earlier investigation of the same model \cite{Stiansen-PRB2012}. The 2D dissipative quantum XY model has a number of applications. Originally proposed \cite{ChakraKivel1} in connection with superconductor to insulator transitions in thin metallic films \cite{Goldman-SIT-rev}, the same model describes the collective fluctuations in the planar itinerant fermion ferromagnet as well as the incommensurate Ising and the commensurate or incommensurate planar antiferromagnet \cite{CMV-PRL2015}, \cite{CMV-IOP2016}. It is also the model for the fluctuations of the loop-current ordered state \cite{simon-cmv} proposed and observed \cite{Bourges-rev} in the underdoped region of the cuprates.

We had focussed in I on the dissipation induced phase transitions in which the phase transition from the disordered state to the (2+1)D quantum ordered state occurs with the spatial correlation length $\xi_r$ proportional to logarithm of the temporal correlation length $\xi_{\tau}$. This may be said to imply a dynamical critical exponent $z \equiv d \ln \xi_{\tau}/d \ln \xi_r = \infty$. It is well understood that without dissipation, the model is Lorentz-invariant, so that  $\xi_{\tau} \propto \xi_r$, i.e. $z=1$. We present here the results of Monte-Carlo calculations to investigate the phase diagram and correlation functions in an extended range of parameters to exhibit the transition of the dynamical critical exponents from $z=1$ to $z = \infty$ as the dissipation is varied. We also investigate further the region of occurrence of a 2D spatial ordered region without temporal order, found in earlier Monte-Carlo calculations \cite{Stiansen-PRB2012, ZhuChenCMV2015}.  We  also discuss the question of the anisotropy of the correlation lengths in time and space in the non-dissipative as well as the dissipative model, as well as the finite temperature cross-over from a 2D Kosterlitz-Thouless type phase to 2+1 D order.

In I, the transition from the quantum disordered to the ordered phase was studied by varying the dissipation parameter $\alpha$, and $\xi_r$ and $\xi_{\tau}$ were calculated as a function of the deviation of the parameter $\alpha$ from its critical value. We complete the work in this region by calculating the correlation lengths near the transition by also varying the parameters for the kinetic energy and the potential energy in the model. We find that the relation $\xi_r \propto \ln  \xi_{\tau}$ is preserved, but that the dependence of $\xi_{\tau}$ on the deviation from critical value of these parameters is an algebraic singularity, rather than an essential singularity found for variation of $\alpha$ from its critical value. These calculations and results have 
important application. While the (2+1)D superconductor-insulator transitions  \cite{Goldman-SIT-rev} are often tuned by varying dissipation through variation of disorder or film thickness, the transitions in metallic anti-ferromagnets are often tuned by varying pressure \cite{HvLRMP2007}. In that case, one expects that the transition is tuned by the variation of the ratio of the kinetic energy to the potential energy.
Often, however, the transition in anti-ferromagnets as well as in the cuprates is tuned by doping.
In that case one expects both the dissipation and the  the ratio of the kinetic energy to the potential energy to be varied. Since the cross-over from quantum-critical to quantum properties occurs as 
$T$ changes from more than to less than $ O(\xi_{\tau}^{-1})$, there should be a vastly wider region of quantum-critical properties for dissipation tuned transitions than for the transitions tuned by the ratio of the kinetic to the potential energy.

Several, but not all, of the Monte-Carlo results have also been obtained in a recent analytical leading order renormalization group calculation \cite{HouCMV2016} on a model of topological excitations derived from the starting dissipative quantum XY model. The transformation of the original model to that of topological excitations is possible only for finite dissipation. Therefore it has not been possible to get analytic results for the transition from the Lorentz-invariant $z=1$ critical point to the $z \to \infty$ critical point at dissipation above a critical value. The present work provides such results. 

 
We also note here the  principal analytical works on the 2D-XY model without dissipation \cite{Doniach1981, MPAFisher86, Chakra2-88, Zimanyi1997}. Our results in the dissipation free model are consistent with these works. We also obtain in our calculations, the rather obvious finite temperature cross-over in that model to the Kosterlitz-Thouless quasi-ordered phase. The crossovers and the transitions in the dissipative model are new and interesting features and may have experimental relevance.

In I and most of the work here, we study the Caldeira-Leggett \cite{CaldeiraLeggett} form of dissipation. In this dissipation, the variable $\theta$ is unbounded. An alternate form of dissipation is for $\theta$ to be a compact variable. In that case, one suspects from Monte-Carlo calculations on a related model \cite{Sperstad2011} that the critical fluctuations remain in the $z=1$ class, i.e. of the same form as without dissipation. We investigate here and find this to be true in the quantum-XY model. Since, both forms of dissipation are allowed for the physical problems of interest, a worthwhile investigation is to study the transition when both forms of dissipation are allowed. We present some results here. We find that over a wide range of the compact dissipation parameter, the transition driven by the Caldeira-Leggett dissipation remains unaffected, i.e. it is of the $z \to \infty$ class.


\section{The Model}
\label{sec:model}

\subsection{(2+1)D Quantum dissipative XY model}
\label{sec:model}

We investigate the action $S$ of  the 2D-DQXY model for the angle $\theta({\bf x}, \tau)$ of fixed-length quantum rotors at space-imaginary time point $({\bf x}, \tau)$:  
\begin{eqnarray}
\label{model}
S &=&-K_0 \sum_{\langle {\bf x, x}' \rangle} \int_0^{\beta} d \tau \cos(\theta_{{\bf x}, \tau} - \theta_{{\bf x}', \tau}) \nonumber \\
& +& \frac 1 {2E_c} \sum_{{\bf x}} \int_0^\beta d \tau \left( \frac{d \theta_{{\bf x}}}{d\tau}\right)^2  \nonumber \\
&+&  \frac{\alpha}{4\pi^2} \sum_{\langle{\bf x, x}'\rangle} \int d \tau  d\tau' \frac {\pi^2}{\beta^2} \frac {\left[(\theta_{{\bf x}, \tau} - \theta_{{\bf x}', \tau})  -(\theta_{{\bf x}, \tau'} - \theta_{{\bf x}', \tau'}) \right]^2}{
\sin^2\left(\frac {\pi |\tau-\tau'|}{\beta}\right)} \nonumber \\
&-& h_4 \sum_{{\bf x}} \int d\tau \cos(4\theta_{{\bf x},\tau}).
\label{eq:model}
\end{eqnarray}
 $\tau$ is periodic in $[0,\beta]$, where $\beta = 1/(k_B T)$.  $\langle {\bf x, x}'\rangle$ denotes nearest neighbors. The first term is the spatial coupling term as in classical XY model. The second term is the kinetic energy where the charging energy $E_c$ serves as the moment of inertia of the rotors. The third term describes quantum dissipations of the ohmic or Caldeira-Leggett type~\cite{CaldeiraLeggett}. 
The last term describes effects of anisotropy on a lattice with four-fold anisotropy.
 
 In the Monte-Carlo calculations, a 2D square lattice with $N\times N$ sites is used. Periodic boundary conditions are imposed along both $x$ and $y$ directions. The imaginary time axis $[0,\beta]$ is spit into $N_{\tau}$ slices. In the discretized (2+1)D lattice,  the action can be written as \cite{Stiansen-PRB2012, ZhuChenCMV2015}:
\begin{eqnarray}
S &=& - K \sum_{\langle{\bf x},{\bf x}'\rangle, \tau} \cos(\Delta\theta_{{\bf x},{\bf x}',\tau}) 
+\frac {K_{\tau} } {2}  \sum_{\bf x, {\tau}}  ( \theta_{{\bf x},\tau }-\theta_{{\bf x},\tau-1})^2  \nonumber \\
&+& \frac{\alpha}{4\pi^2}  \sum_{\langle{\bf x},{\bf x}'\rangle, \tau, \tau'} \frac {\pi^2}{ N_{\tau}^2} \frac {[\Delta\theta_{{\bf x},{\bf x}',\tau}-\Delta\theta_{{\bf x},{\bf x}',\tau'}]^2}{
\sin^2\left(\frac {\pi |\tau-\tau'|}{N_{\tau}}\right)} ),
\label{eq:modeld}
\end{eqnarray}
where $K_{\tau} \equiv 1/(E_c\Delta \tau)$, $K\equiv K_0\Delta \tau$ are dimensionless kinetic energy and potential energy parameters. They have been normalized by the ultra-violet cut-off $\Delta \tau = \beta/N_{\tau}$. In this representation, the temperature is controlled by $N_{\tau}^{-1}$. The calculations are asymptotically correct for the quantum problem where $1/\beta = T \to 0$ and $N_{\tau} \to \infty$, with their product held constant. This requires in practice that we ensure that the results converge in the range of $N_{\tau}$ actually studied.  

The four-fold lattice anisotropy $h_4$ is marginally irrelevant in the classical XY model \cite{JKKN1977}, and irrelevant at the quantum transition \cite{Aji-V-qcf1, Aji-V-qcf2}, as verified in I. We have not included it in the calculation in this paper. 

Quite obviously, a larger potential energy parameter $K$ favors ordering and a larger kinetic energy parameter $K_{\tau}$ prefers a quantum-disordered state. In the absence of dissipation, i.e. for $\alpha = 0$, the transition to a 2+1 D ordered state at $T=0$ therefore occurs for $1/(KK_{\tau})$ below a critical value. In such a Lorentz invariant model, the ratio of the kinetic to the potential energy determines the effective velocity of the critical modes, and therefore also the ratio of the correlation length in time with respect to that in space.  In interpreting the Monte-carlo results, it is useful to note that $1/(KK_{\tau})$ is independent of $T$ and $N_{\tau}$. On the other hand, $K_{\tau}/K = (N_{\tau}T)^2/(K_0E_c)$. Both the effects of the physical anisotropy parameter $K_0E_c$, as well as information about finite temperature cross-over due to the inevitable finite $N_{\tau}$ in the calculations are given by $K_{\tau}/K$. These points are important to bear in mind when reading the phase diagrams deduced below.

The dimensionless dissipation parameter is defined as $\alpha=R_Q/R_s$, where $R_s$ is the normal resistivity and  $R_Q= h/4e^2$ is the quantum of resistivity. Please note that $\alpha/(4\pi^2)$ defined in this paper corresponds to $\alpha$ in Refs. \onlinecite{Stiansen-PRB2012, ZhuChenCMV2015}. We have made this change so that the critical point in the model as a function of dissipation occurs near the $\alpha =1$ defined in this paper. Also, with this definition, the magnitude of  $\alpha$ is easily interpreted physically.

 A larger $\alpha$ (smaller normal state resistivity) promotes the ordered state.  Since in two dimensions, strong localization occurs at low temperatures if the normal state resistance is of the order of the quantum of resistance, it can be argued that only models with $\alpha \gtrsim 0.2$ are relevant for applications to experiments for any of the metallic systems for which the present work may be applicable. 

The procedure, accuracy and limits of validity of the Monte-Carlo calculations have all been given earlier in I in Sec. 2C. The same apply for the investigation here and need not be discussed again. In I, several physical quantities were defined to characterize the phases and the correlations. For convenience, they are defined again here: \\
{\it Action susceptibility.} The action susceptibility is defined as 
\begin{equation}
\chi_S = {1\over N^2 N_{\tau} } \left(\left\langle S^2\right\rangle -\left\langle S \right\rangle ^2\right),  
\end{equation} 
where $\langle \ldots \rangle$ denotes averaging over the $O(10^6)$ Monte-Carlo measurements. 
In classical systems, as $S=\beta H$, $\chi_S$ is related to the specific heat, $\chi_S = C_V/k_B$.  At $T \to 0$, it is a measure of zero-point fluctuations which are expected to be singular at the critical point due to the degeneracy in the spectra. 

{\it Helicity Modulus.}  The helicity modulus or spatial stiffness is defined as the change of energy due to
a slow twist of spins along the spatial direction, or
\begin{eqnarray}
\Upsilon_x &=& {1 \over N^2 N_{\tau}} \left\langle \sum_{\langle{\bf x},{\bf x}'\rangle} \sum_{\tau} \cos(\Delta \theta_{{\bf x},{\bf x}', \tau}) \right\rangle  
  \nonumber \\
  &-& {K \over N^2 N_{\tau}} \left\langle \left( \sum_{\langle{\bf x},{\bf x}'\rangle} \sum_{\tau} \sin(\Delta \theta_{{\bf x},{\bf x}', \tau}) \right)^2 \right\rangle .
\end{eqnarray}
In the disordered state, the two terms have comparable contributions and $\Upsilon_x \to 0$. In an ordered phase, the second term vanishes, so that $\Upsilon_x$ is finite. 

{\it Order parameter.} For XY model, the order parameter ${\bf M} ({\bf x},\tau) = (\cos \theta_{{\bf x},\tau}, \sin \theta_{{\bf x},\tau})$. Its modulus, the magnetization in the plane, is defined as 
\begin{equation} 
M = \frac {1}{N^2N_{\tau}} \left\langle \left| \sum_{{\bf x},\tau}  e^{i \theta_{{\bf x}, \tau}} \right| \right\rangle.
\end{equation}
In classical 2D XY model, the ordered phase has a quasi long-range order, where $M \sim (1/N)^{1/(8\pi K)}$, vanishes for $N \to \infty$.   We also found it illuminating to calculate $M_{2D}$, the magnitude of magnetization in the planes at a given time $\tau$ and then average it over the $\tau$. This is equivalent to finding the Kosterlitz-Thouless order parameter at each time slice and then averaging over the time slices. 
\begin{equation} 
M_{2D} = \frac {1}{N^2N_{\tau}} \left\langle \sum_{\tau} \left| \sum_{{\bf x}}  e^{i \theta_{{\bf x}, \tau}} \right| \right\rangle.
\end{equation}
By definition $M \leqslant M_{2D}$. Also $M=M_{2D} \ne 0$ (for $N \to \infty$) only if there is perfect long range order across time as well as space. 

{\it Local mean-square fluctuations of $\theta$ in time}. We also calculate the spatially local mean-square fluctuations in $\theta(\tau)$, denoted by $W_{\theta}^2$ :
\be
W_{\theta}^2 \equiv \frac{1}{N_{\tau}}\Big\langle \sum_1^{N_{\tau}}\big(\Delta\theta_{\tau} -\overline{\Delta\theta}\big)^2\Big\rangle.
\ee
Here $\overline{\Delta\theta} = \frac{1}{N_{\tau}}\sum_1^{N_{\tau}} \Delta\theta_{\tau}.$
As shown in Ref. \cite{Stiansen-PRB2012} and in I, a rapid change in $W_{\theta}^2$ is 
a measure of critical slowing down in transitions.

{\it Correlation Function of the Order Parameter.} The principal results for the quantum-critical fluctuations are given by the order parameter correlation functions:
\begin{equation}
G_{\theta} ({\bf x}, \tau) = \frac {1}{N^2N_{\tau}} \sum_{{\bf x}',\tau'}\left\langle e^{i (\theta_{{\bf x'}+{\bf x}, \tau'+\tau} - \theta_{{\bf x}', \tau'})}\right\rangle.
\end{equation}
$G_{\theta} ({\bf x} \to \infty, \tau \to \infty) \to M^2$ while $G_{\theta} ({\bf x} \to \infty, \tau =0) \to M_{2D}^2$.

\begin{figure}[tbh]
\centering
\includegraphics[width=0.48\columnwidth]{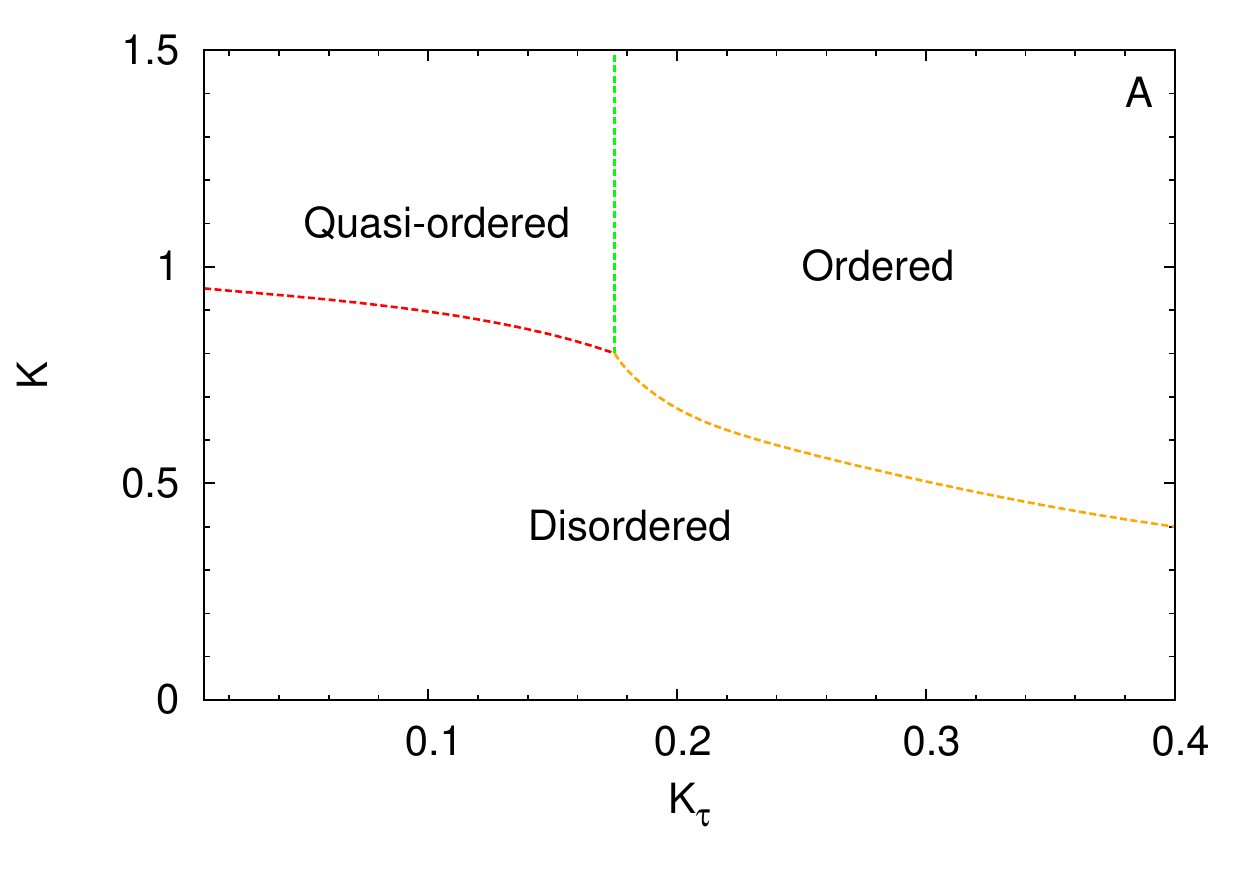}
\includegraphics[width=0.48\columnwidth]{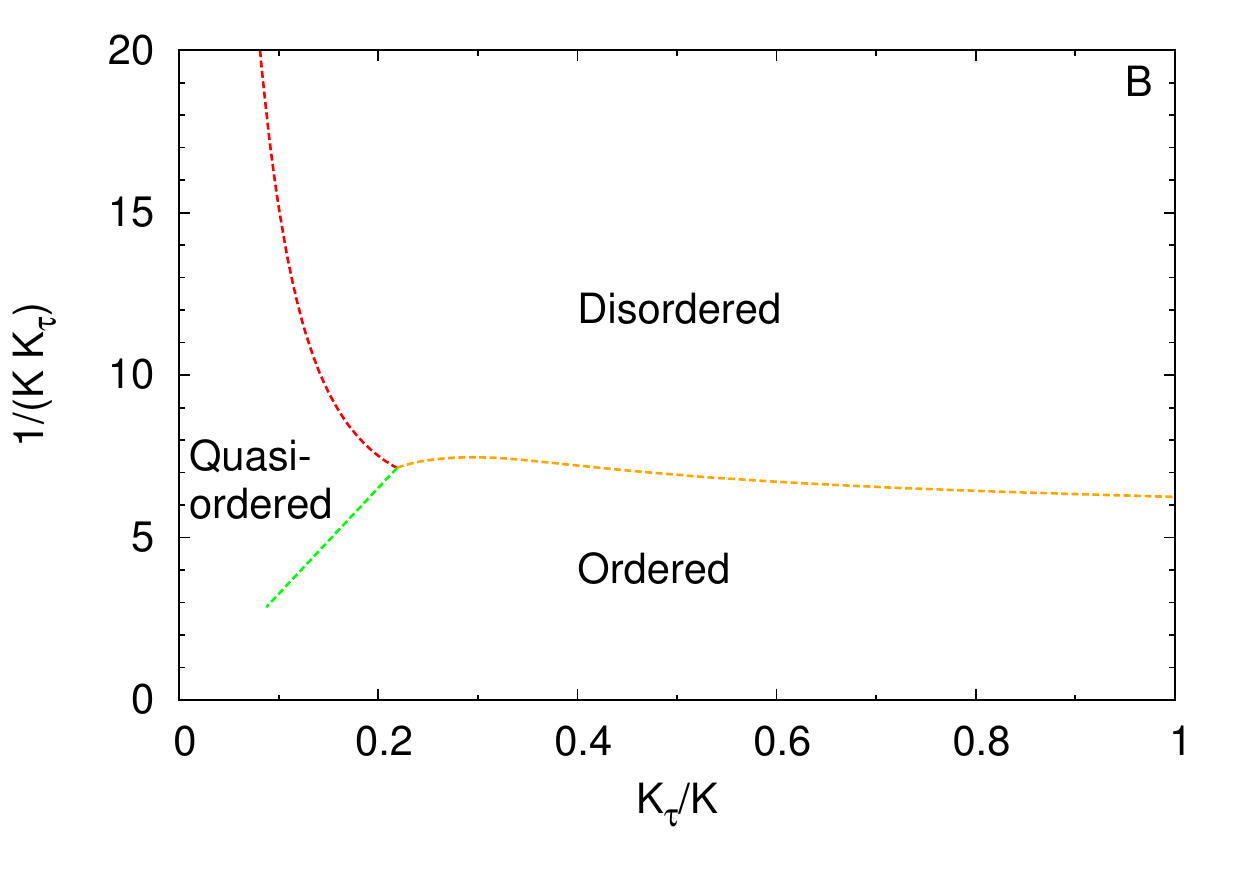} 
\caption{A: Phase diagram for $\alpha=0$ in the $K_{\tau}-K$ plane, and B: the same in the $(K_{\tau}/K)-1/(K_{\tau}K)$ plane. This phase diagram is constructed from calculations of various physical quantities in Figs. \ref{fig:statica0dtoqo} and \ref{fig:statica0qo2o} for the the values of $K_{\tau}$ and $K$ for which the results are presented there. As explained in the text, $(K_{\tau}K)$ is independent of temperature, while $1/(K_{\tau}/K)$ depends on temperature as well as on the ultraviolet spatial and temporal scales in the Monte-Carlo simulations.  At $T\to 0$, the transition from the Disordered to the Kosterlitz-Thouless type Quasi-ordered phase as well as from the latter to the (2+1)D Ordered phase occur as cross-overs. The transition from the Disordered to the Ordered phase is of the 3D classical XY universality class. The implications of part B of the diagram for finite $T$ cross-overs and effects of temporal and spatial anisotropy are discussed in the text.}
\label{Fig-phdia-a=0}
\end{figure}

\section{Phase diagram for $\alpha=0$}

The phase diagram with $\alpha=0$, determined from the calculations described below, is drawn with two different choice of axes, in Fig. (\ref{Fig-phdia-a=0}).  Part B of this figure is more revealing of the cross-overs from finite temperatures to $T\to 0$ and of the effects of anisotropy in the temporal and spatial correlation lengths. The limit of $K_{\tau} \to 0$ is non-analytic, with a cross-over line separating a 2+1 D ordered phase from a region of pronounced correlations of the 2 D quasi-ordered phase of the Kosterlitz-Thouless type with temporal disorder starting from the origin in Fig. (\ref{Fig-phdia-a=0}-B). The  line between the two phases represents the growing correlation of warps (instantons of monopoles-antimonopoles with total charge 0 \cite{Aji-V-qcf3}, fully discussed in I) in the temporal direction.  It shows, as expected, that a phase with 2D quasi-order exists for $1/E_c \lesssim N_{\tau}T$, (with a finite $N_{\tau}$) where quantum effects are unimportant. The line ends for $1/(KK_{\tau}) \approx 7$ separating the disordered region from the 2+1 D ordered region. 

\begin{figure}[tbh]
\centering
\includegraphics[width=0.7\columnwidth]{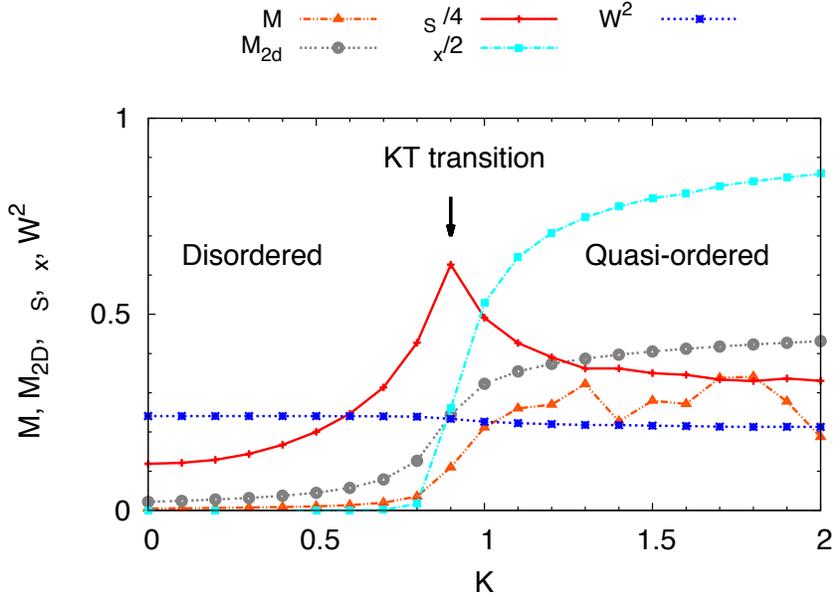}
\caption{Static properties defined in the text, for $\alpha=0$, $K_{\tau}=0.1$ and varying $K$ over the 21 points shown. The results shown are for $N = 50$ and $N_{\tau} =100$. A rapid growth and fall in $\chi_S$ denotes the passage across a symmetry breaking for a quantum transition. The rapid growth of the helicity modulus $\Upsilon_x$ and $M$ and $M_{2D}$, as discussed in the text, reveal a finite size crossover from the disordered phase to quasi-ordered phase. The smooth behavior of $W^2_{\theta}$  shows that temporal correlations do not change across this crossover, unlike the spatial correlations.} 
\label{fig:statica0dtoqo}
\end{figure}
\begin{figure}[h]
\centering
\includegraphics[width=0.7\columnwidth]{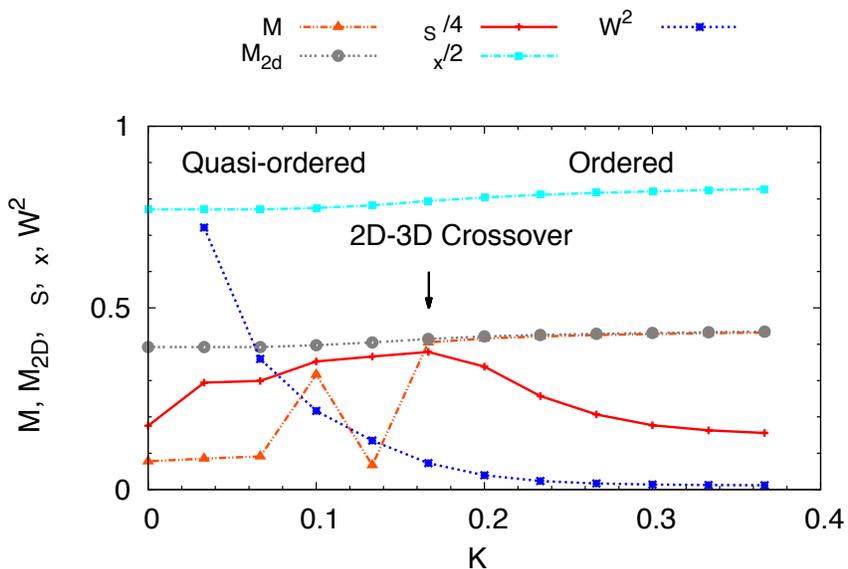}
\caption{Calculation of the specified static quantities showing results consistent with a transition (as $T \to 0$) from the quasi-ordered phase to the ordered phase for $\alpha=0$, $K=1.4$ by varying $K_{\tau}$ over the 12 different values shown. The rather large noise in the data is due to finite size effects discussed in the text.}
\label{fig:statica0qo2o}
\end{figure}
\begin{figure*}[t]
\centering
\includegraphics[width=0.6\textwidth]{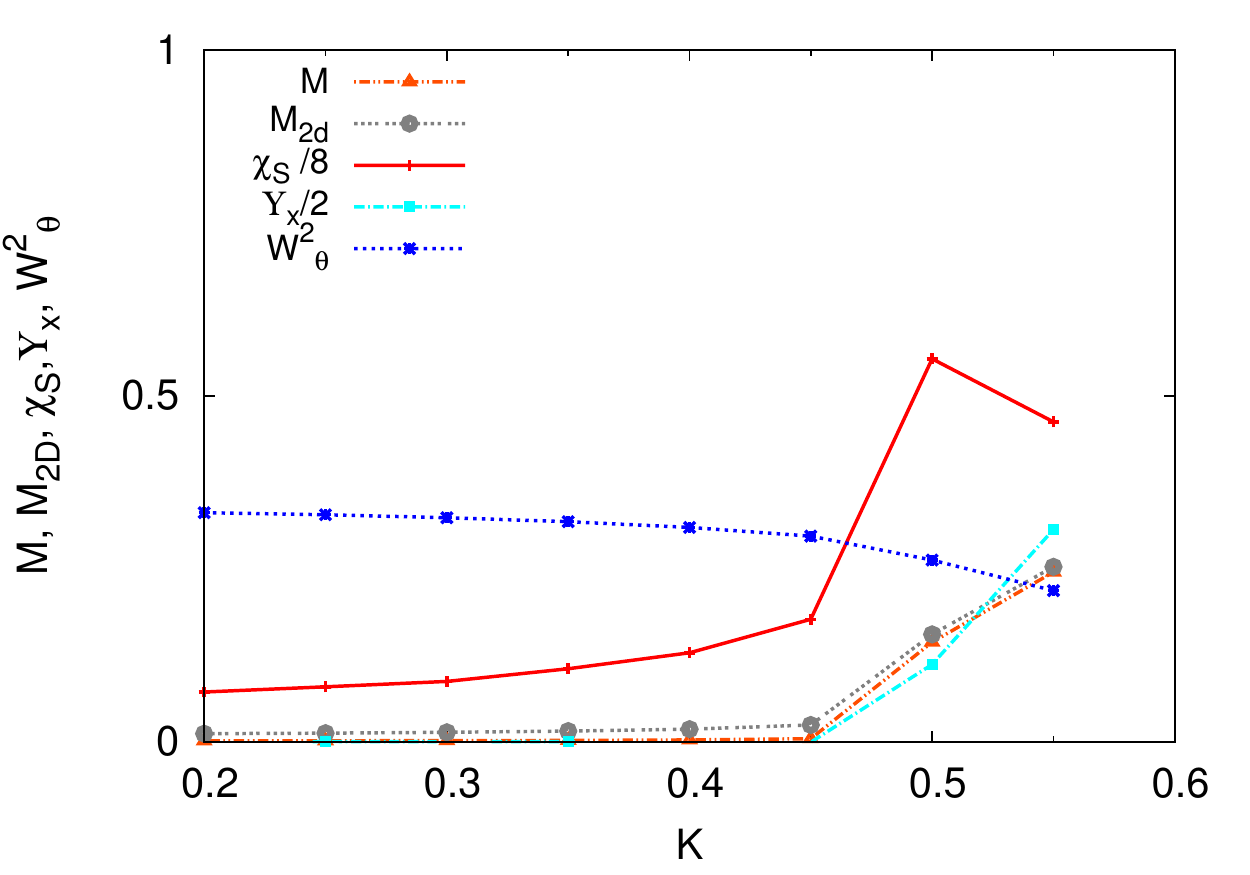}
\includegraphics[width=0.45\textwidth]{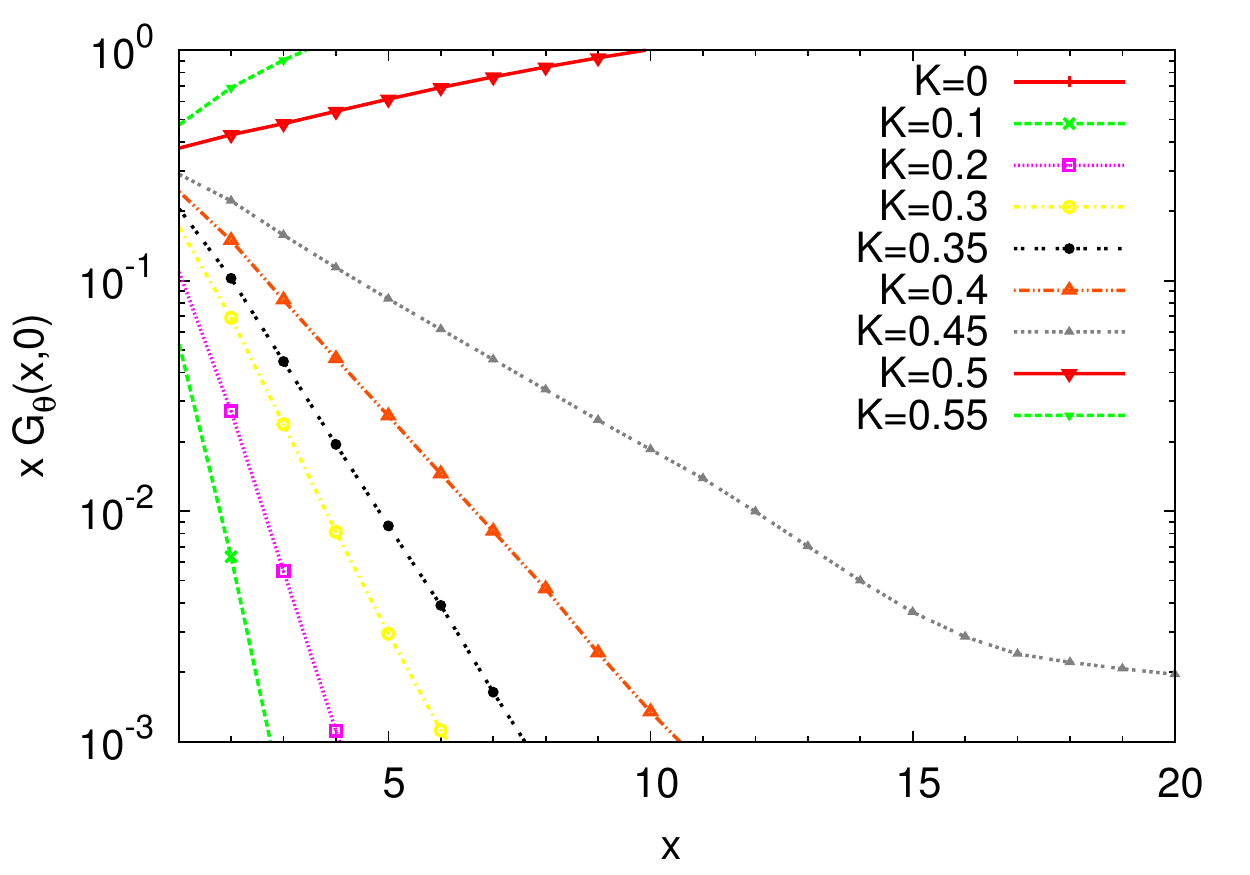}
\includegraphics[width=0.45\textwidth]{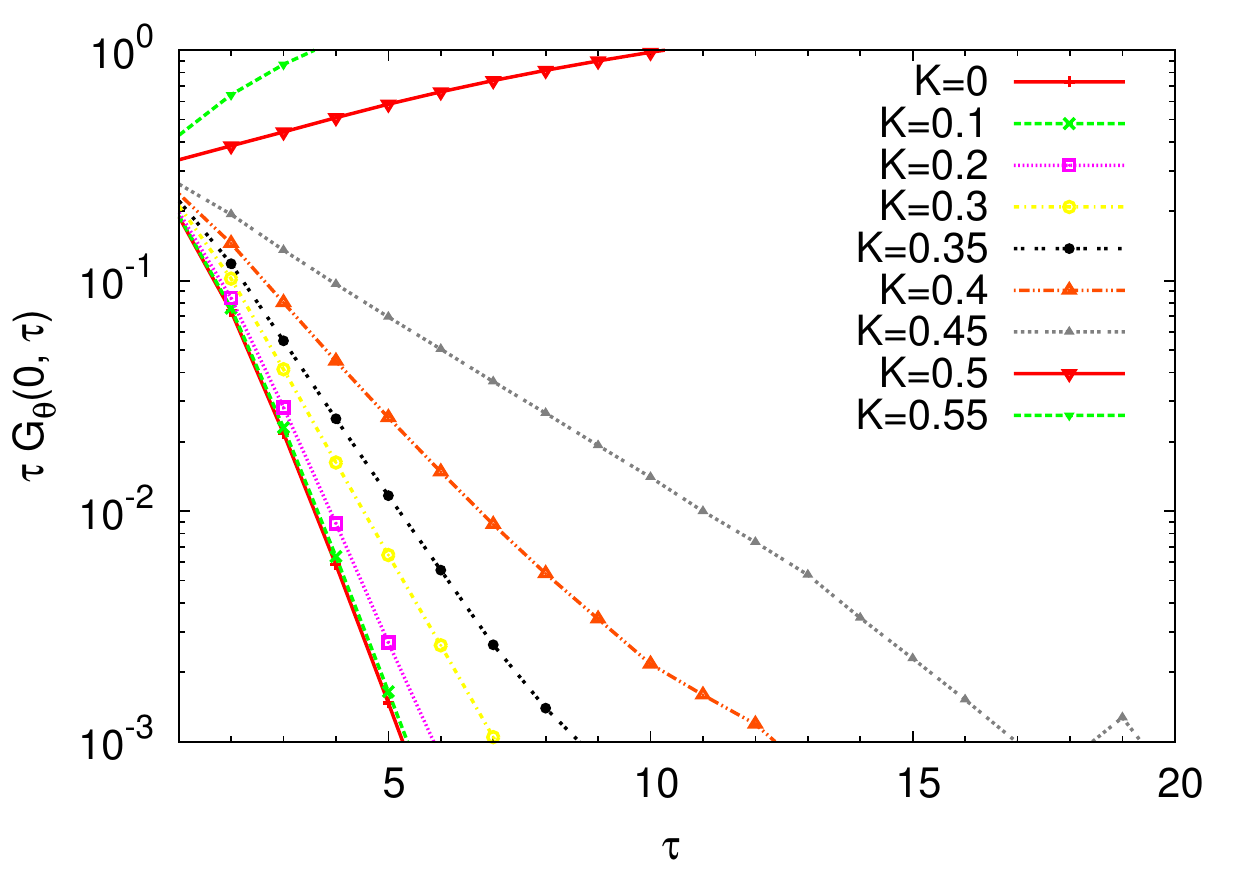}
\includegraphics[width=0.45\textwidth]{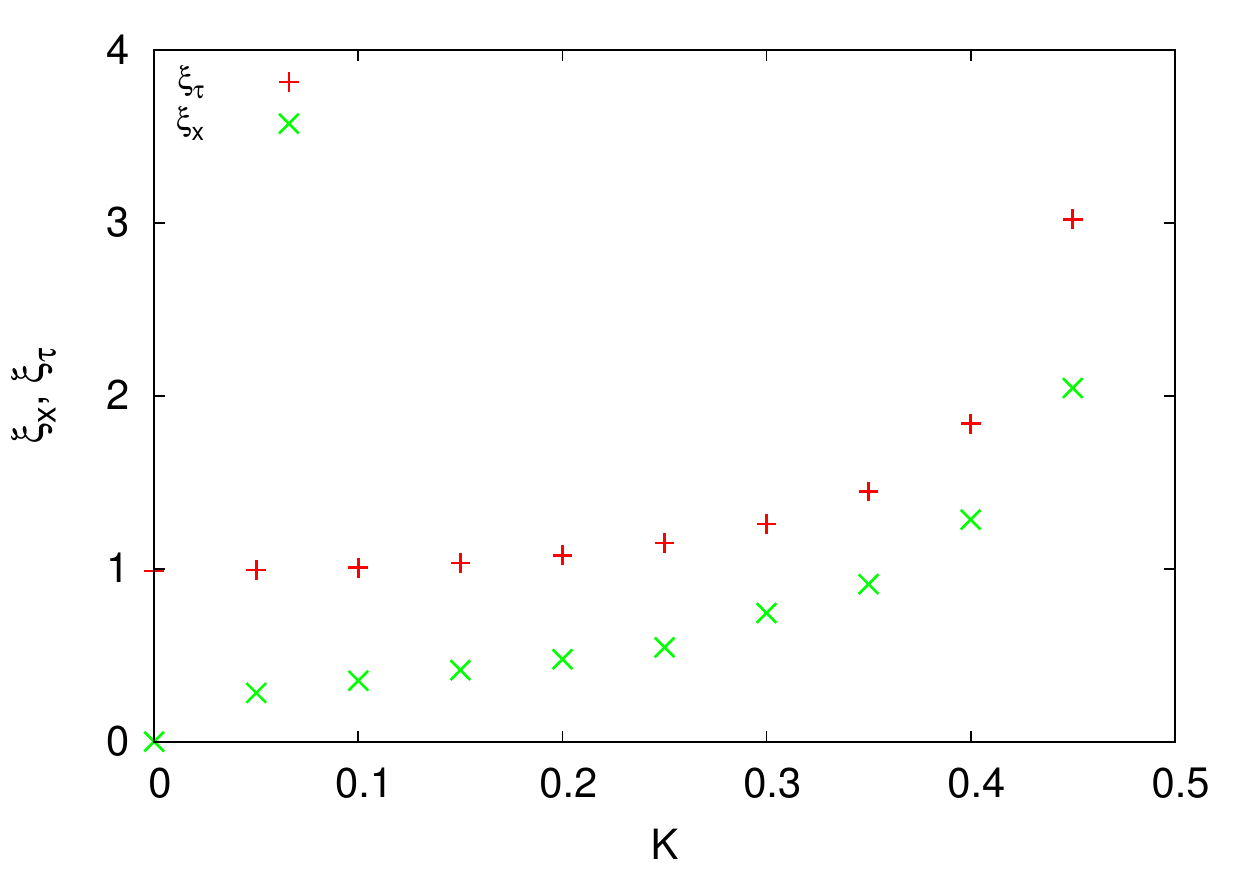}
\includegraphics[width=0.45\textwidth]{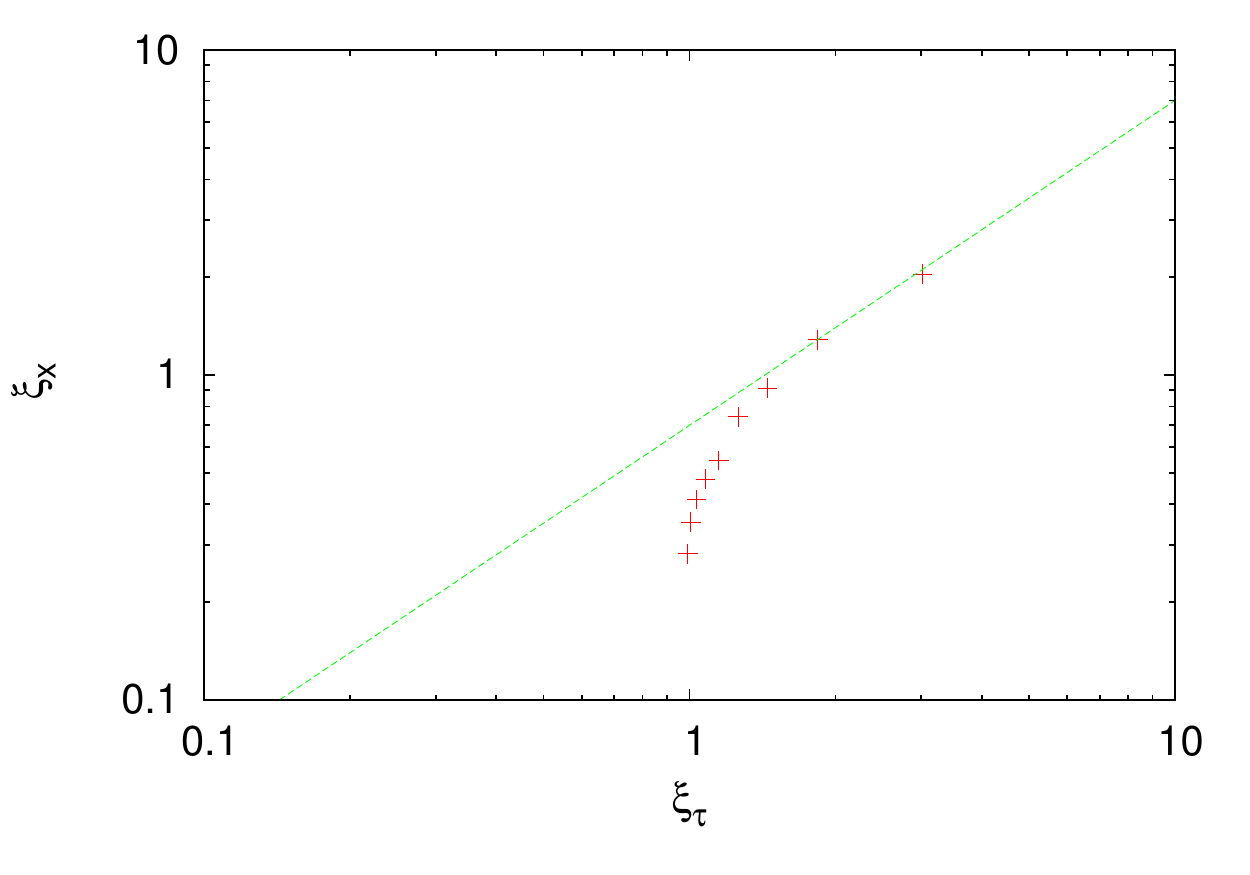}
\caption{A transition from disordered phase to ordered phase when $\alpha=0$. Here $K_{\tau}=0.3$ and $K$ is varied. The critical value for the potential energy parameter, $K_c$ is between 0.45 and 0.5. In the first panel, static quantities are shown: $M, M_{2D}, \Upsilon_x$ achieve non-zero values across the transition. The next panels show the order parameter correlations, for fixed time as a function of $x$ and for fixed $x$ as a function of $\tau$.  We find $G_\theta(x, 0) \sim (1/x)\exp(-x/\xi_x)$ and $G_\theta(0, \tau) \sim (1/\tau) \exp(-\tau/\xi_\tau)$. The final panels shows that close to the transition, the spatial correlation length $\xi_x \sim \xi_\tau$, the temporal correlation length for $\xi_x \gg 1$, i.e. the dynamical critical exponent $z\approx 1$,  consistent with a 3D XY transition.}
\label{fig:a0d2o}
\end{figure*}

The similarity of the phase diagram Fig. (\ref{Fig-phdia-a=0}-A) in the $K-K_{\tau}$ plane to that in the $K-\alpha$ plane for fixed $K_{\tau}$, given in Fig. (2) of I, is illustrative of the fact that reduced kinetic energy promotes a transition to the ordered phase, just as increasing dissipation does. For $\alpha =0$ and $T=0$, there is only one dimensionless parameter which determines the condition for the 2+1 D transition. The purely 2D phase (the quasi-ordered or Kosterlitz-Thouless phase) exists only at finite T but cross-overs to it is exhibited at $T \to 0$, as demarcated by the green line in Fig. (\ref{Fig-phdia-a=0}-A). The situation changes when $\alpha \ne 0$, where we found in I that, within the accuracy of Monte-Carlo determination of the Helicity Modulus $\Upsilon_x$, 2D order parameter $M_{2D}$ and (2+1) D order parameter $M$, a Kosterlitz-Thouless type transition occurs even for $T \to 0$, while the temporal correlations do not change.


\begin{figure*}[tbh]
\centering
\includegraphics[width=0.6\textwidth]{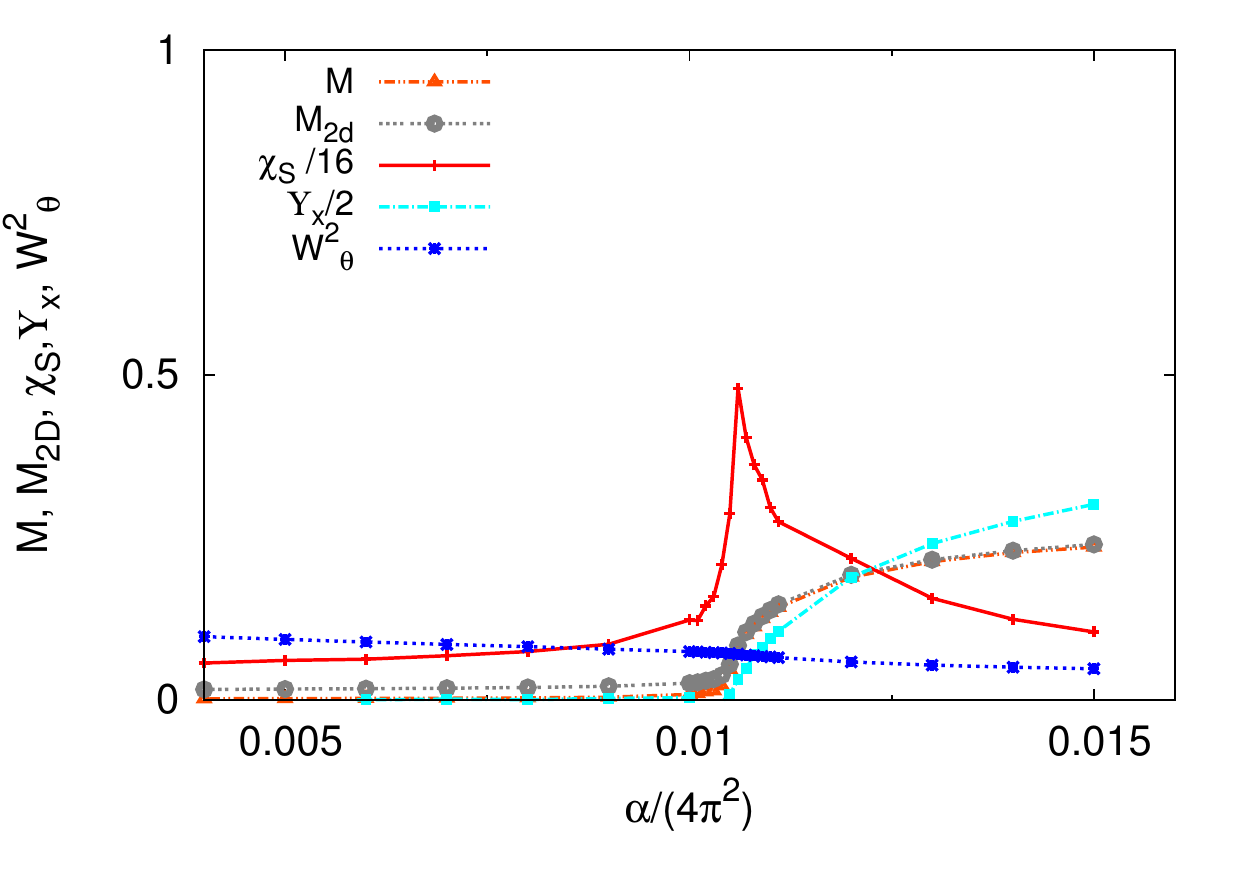}
\includegraphics[width=0.45\textwidth]{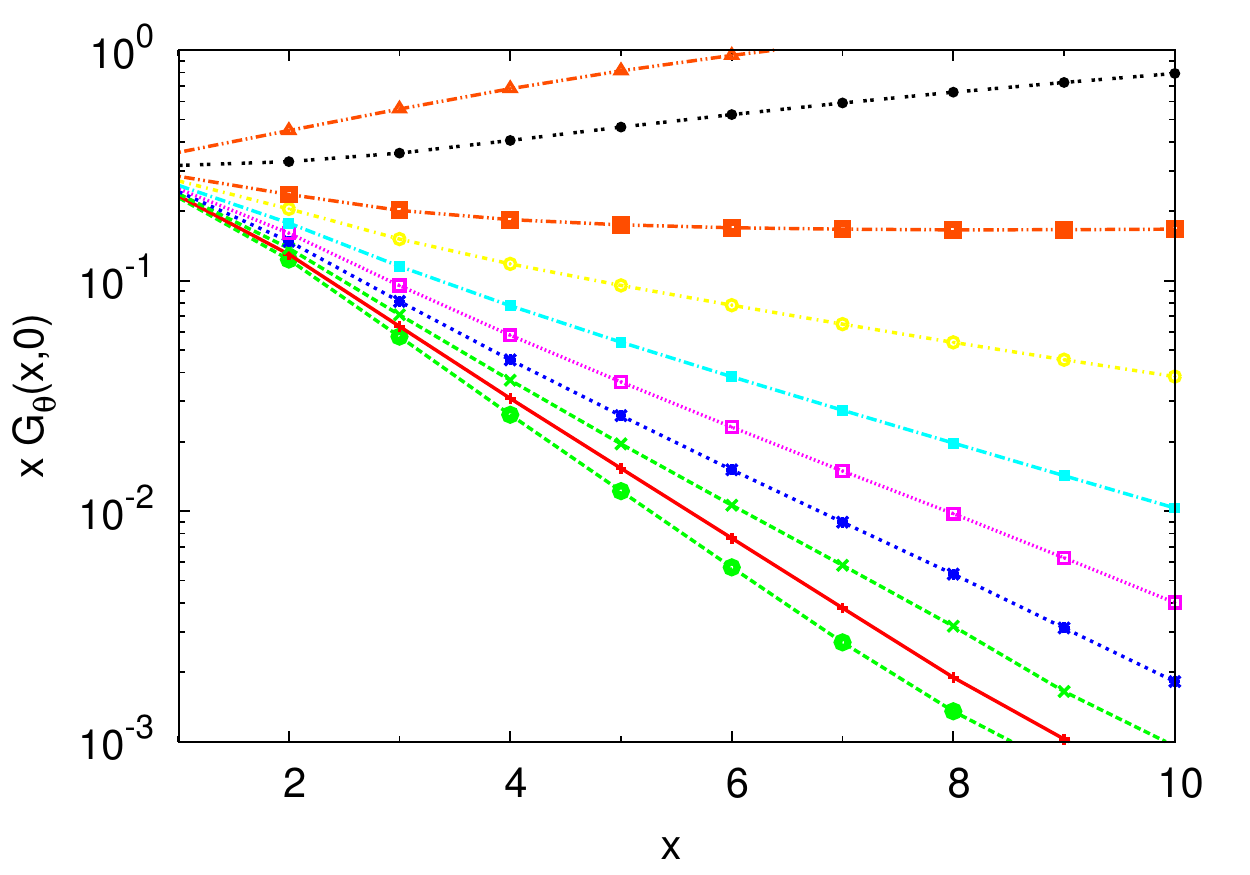}
\includegraphics[width=0.45\textwidth]{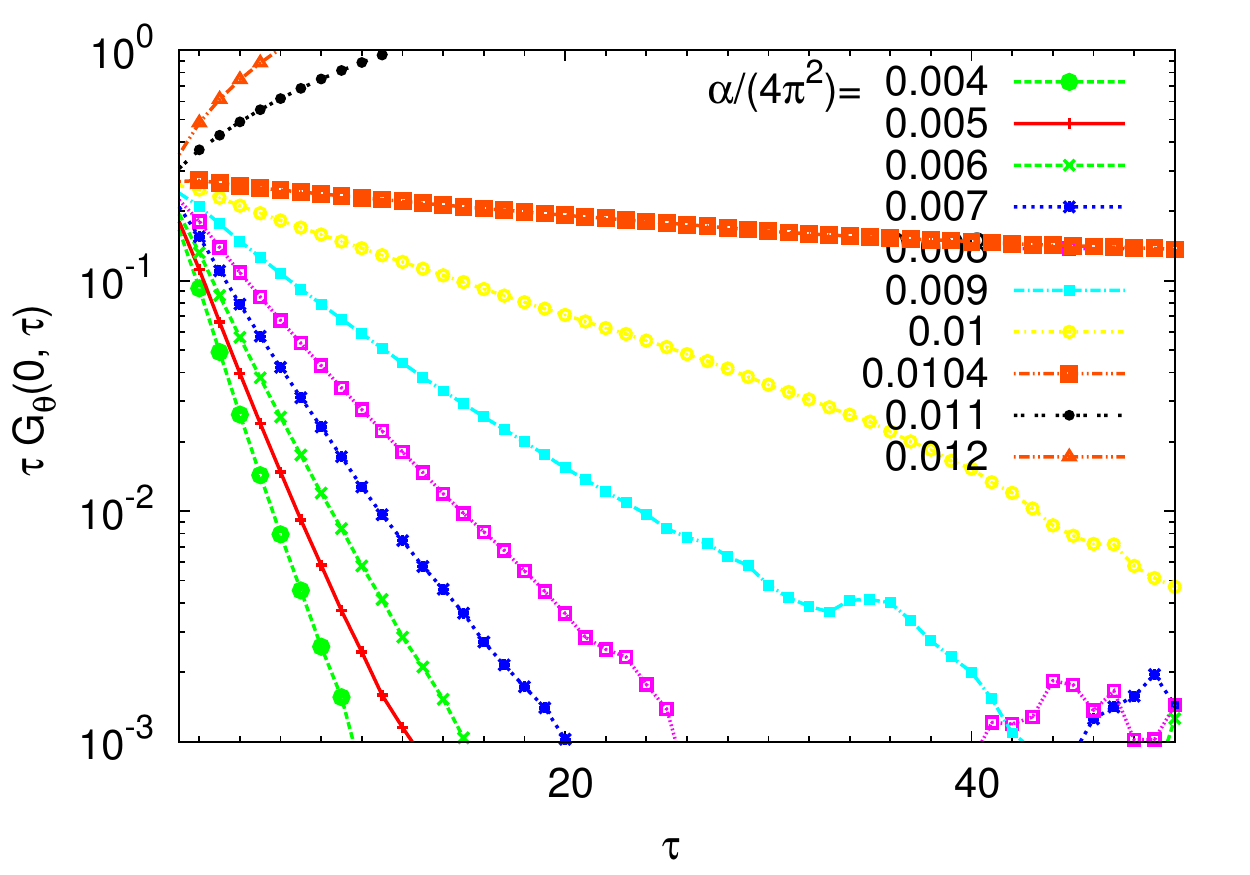}
\includegraphics[width=0.45\textwidth]{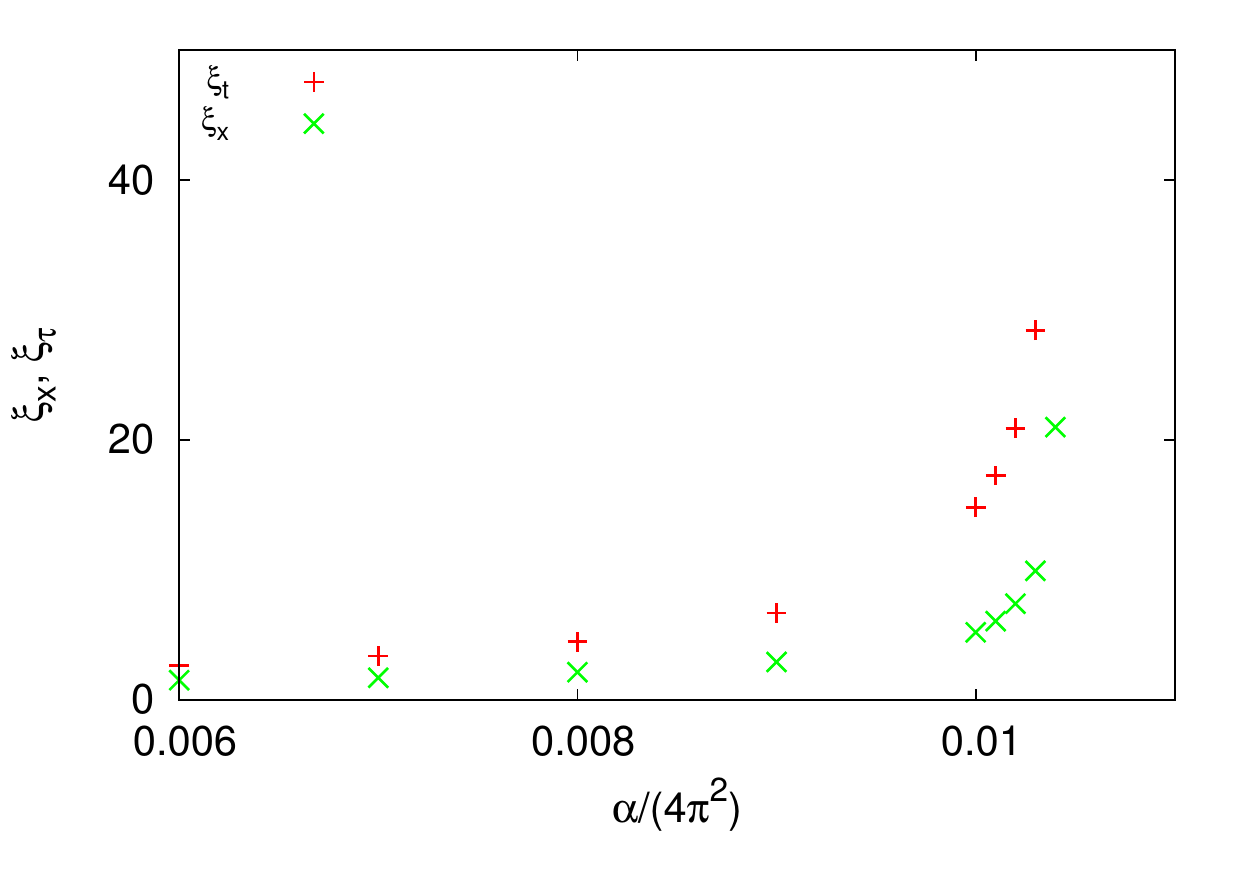}
\includegraphics[width=0.45\textwidth]{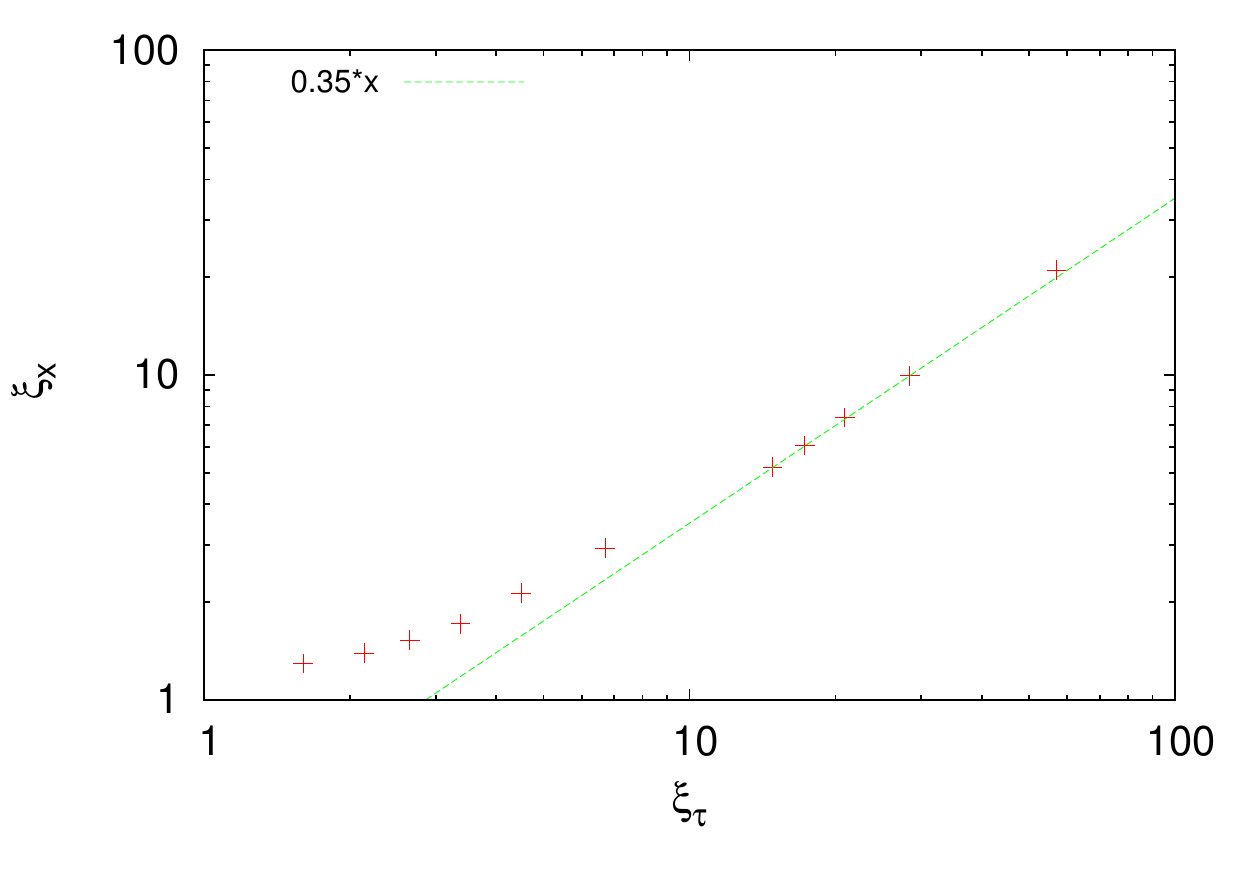}
\caption{This figure shows the same quantities as in Fig. (\ref{fig:a0d2o}) and in the same order but by varying $\alpha$ and for $K_{\tau}=0.2$, $K=0.4$. We find similar results as in the calculations for $\alpha=0$ by varying $K$ or $K_{\tau}$. Most importantly, in the calculations shown in this and the preceding figure, $\xi_x \sim \xi_\tau$ near critical point, i.e. the dynamical critical exponent $z\approx 1$. }
\label{fig:K04Kt02}
\end{figure*}

\begin{figure*}[h]
\centering
\includegraphics[width=0.6\textwidth]{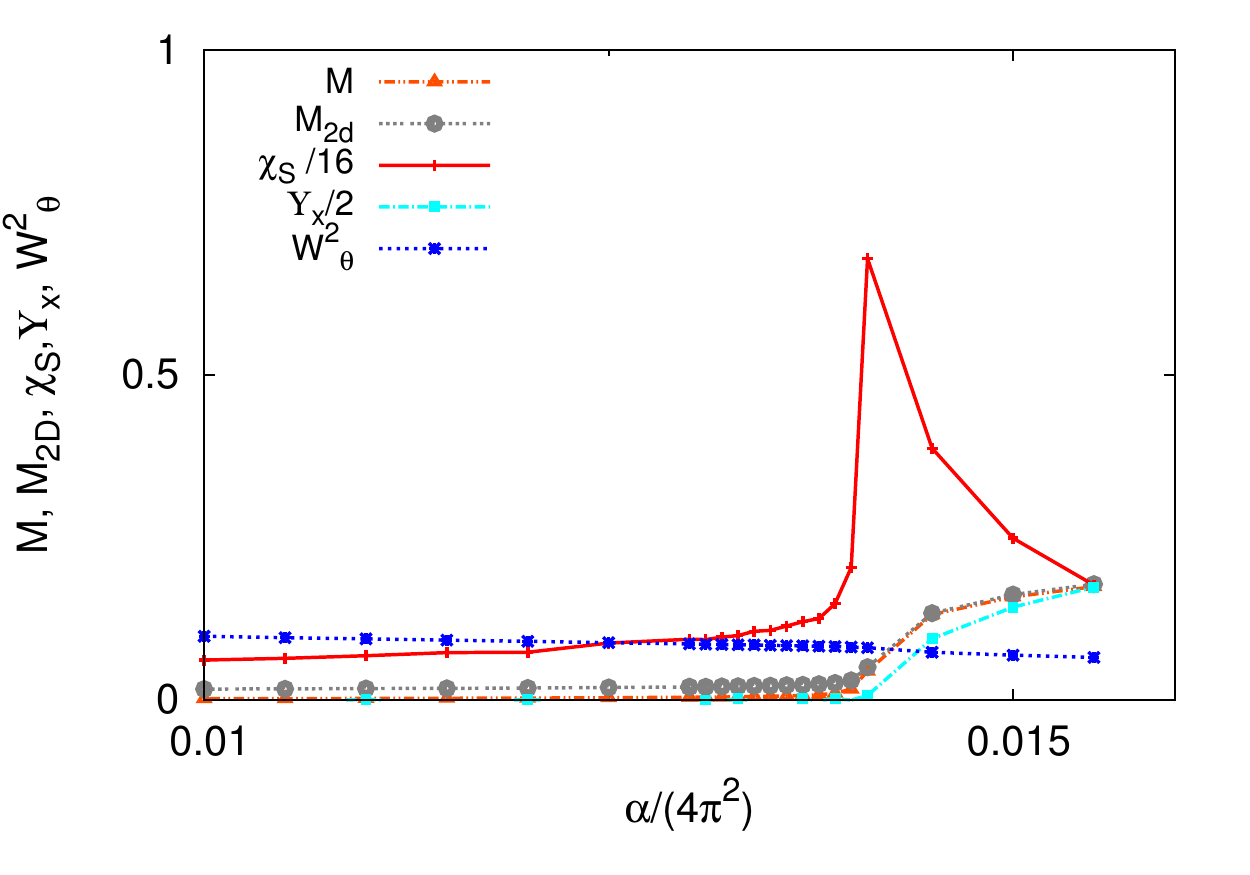}
\includegraphics[width=0.45\textwidth]{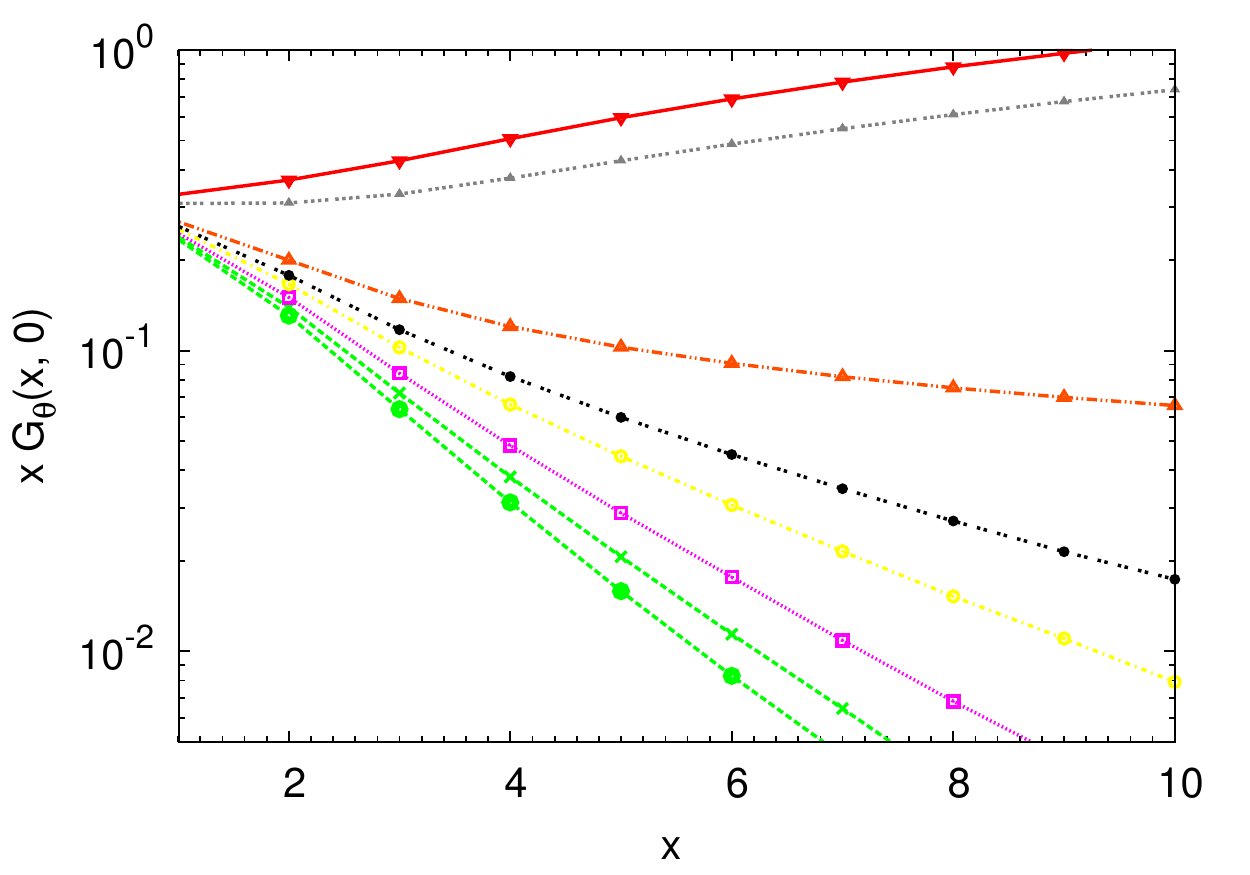}
\includegraphics[width=0.45\textwidth]{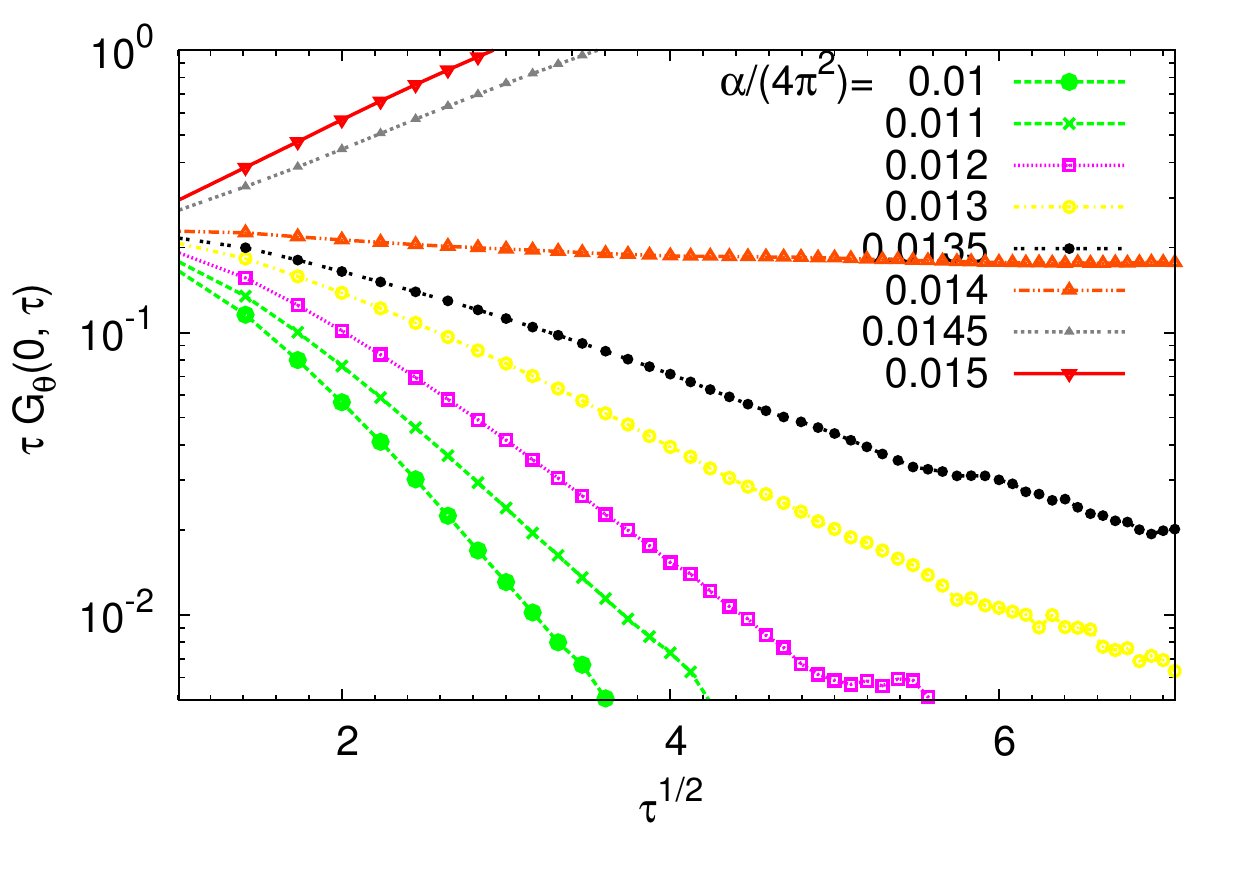}
\includegraphics[width=0.45\textwidth]{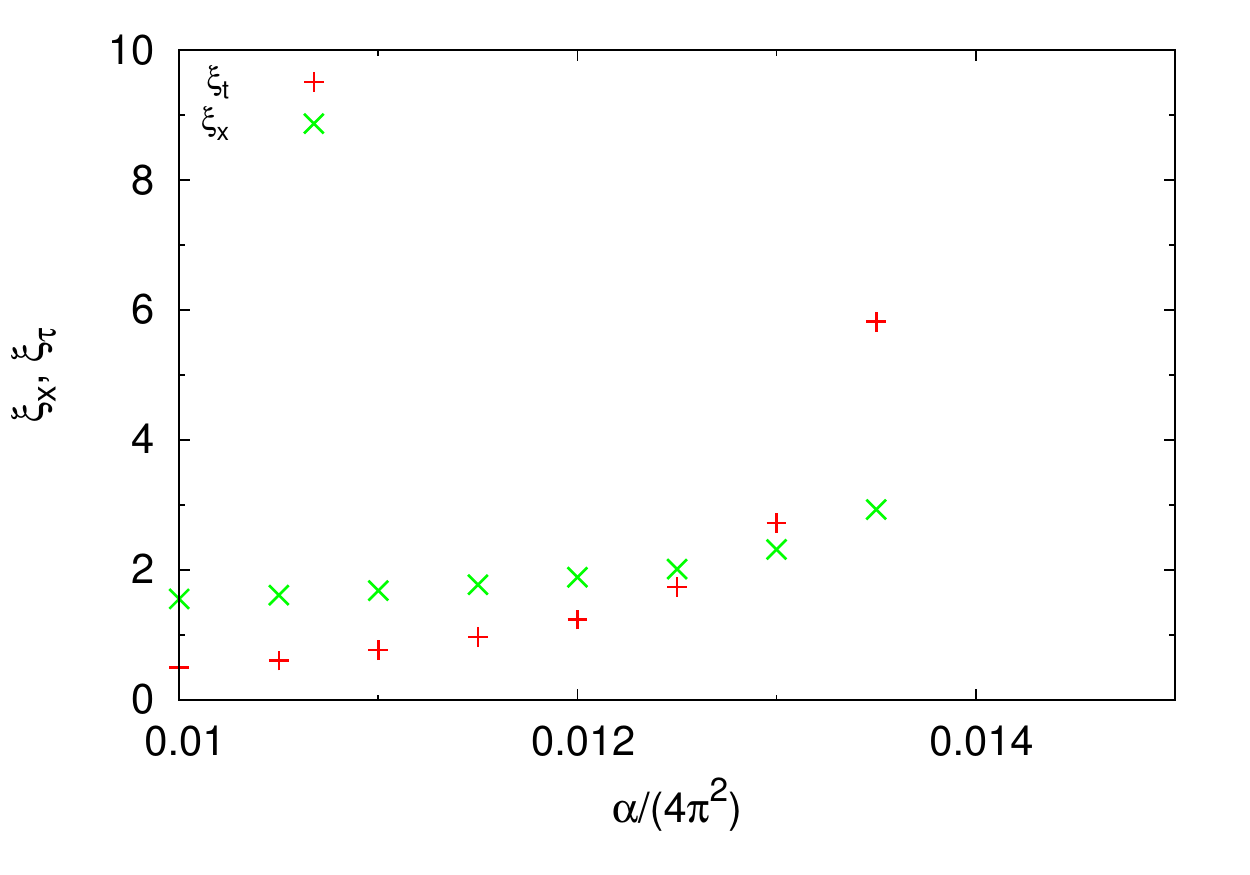}
\includegraphics[width=0.45\textwidth]{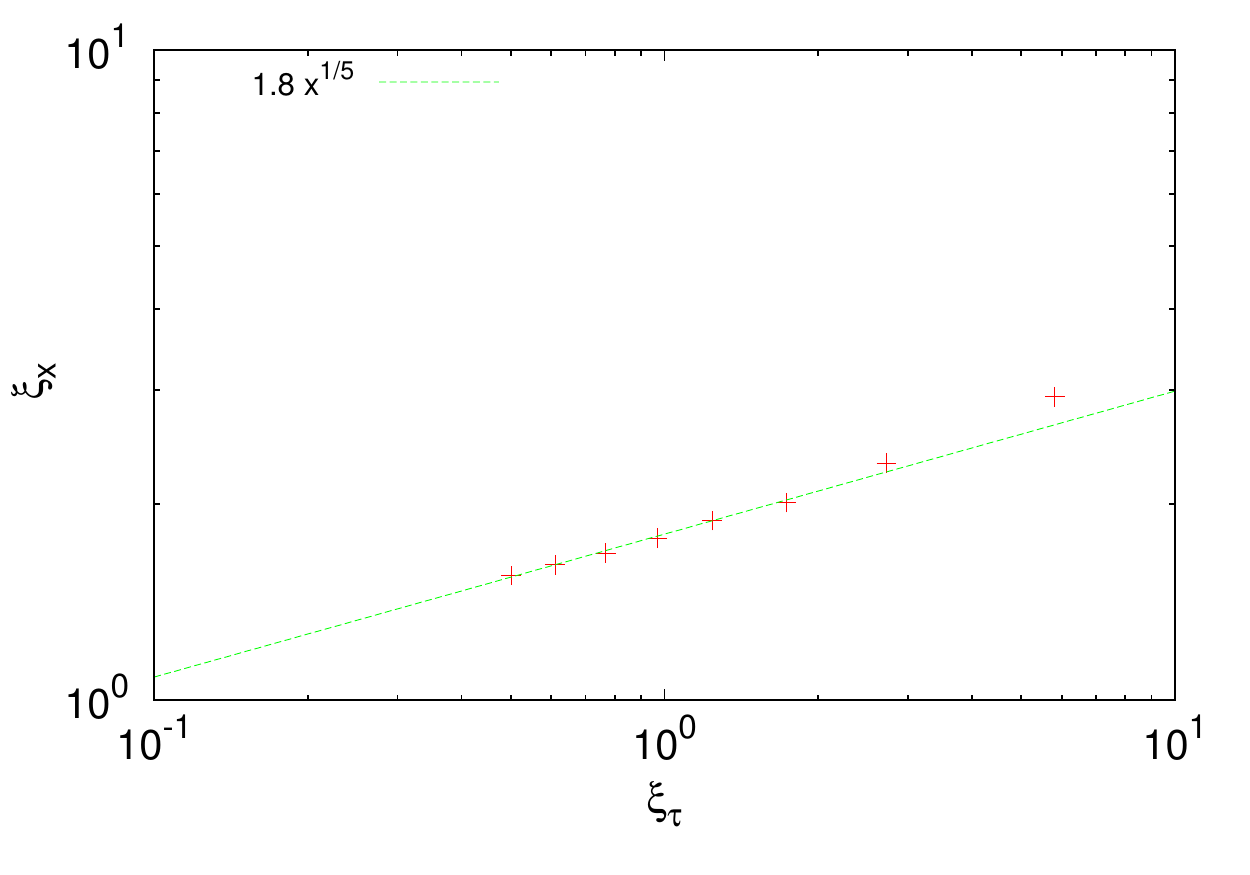}
\caption{This figure shows the same quantities and in the same order as in Figs. (\ref{fig:a0d2o}) and (\ref{fig:K04Kt02}) but with $K_{\tau}=0.15$, $K=0.4$ and $\alpha$ is varied.  A transition from disordered phase to ordered phase is observed above a critical $\alpha$, as obvious from the static quantities on the top-most panel, and consistent with the correlation functions shown in the next pair of panels. But in contrast to the results in the two preceding figures for small $\alpha$, if we fit the relation between correlation length and correlation time by a power law, we find $\xi_x \sim \xi_\tau^{1/z}$, with $z \approx 6$. The relation $\xi_x \sim \log \xi_{\tau}$ yields an equally good fit in the range shown, consistent with $z \to \infty$.}
\label{fig:K04Kt015}
\end{figure*}

We show now results of some of the calculations leading to the phase diagrams in Fig. (\ref{Fig-phdia-a=0}). Fig.(\ref{fig:statica0dtoqo}) shows an example of the disordered to quasi-ordered 2D transition with $\alpha=0$, $K_{\tau}=0.1$ by varying $K$.  Various static properties are plotted. The action susceptibility  $\chi_S$ develops a peak at $K\approx 0.85$. Around this point, the  2 D spatial order parameter $M_{2d}$ as well as the helicity modulous $\Upsilon_x$ show a rapid growth and remain non-zero for further increase in $K$. The mean-square temporal fluctuation in angles, $W^2_{\theta}$ hardly changes showing that there is no growth of correlations in the time-direction. $M$ is non-zero but non-monotonic for $K \gtrsim 0.85$ for the calculation shown which has $N_{\tau} = 100$ but we have obtained results showing that $M$ decreases for larger $N_{\tau}$, and is consistent with being 0 asymptotically. It is difficult to study this behavior systematically as it shows large fluctuations. The results are consistent with what one expects of the quasi-ordered phase with spatial order (in each time slice) as evidence by the $M_{2d}$ in Fig. (\ref{fig:statica0dtoqo}), but disordered along the time direction. This is further corroborated by the study of the  correlation functions.  

 Fig. \ref{fig:statica0qo2o} shows results with $\alpha=0$, $K=1.4$ while  $K_{\tau}$ is varied. Increasing $K_{\tau}$ drives the quasi-ordered 2D phase to the (2+1)D ordered phase. When $K_{\tau} \gtrsim 0.18$, $M \approx M_{2D}$, and $W^2_{\theta}\to 0$ showing rapidly diminishing temporal fluctuations. The purely spatial characteristics such as $M_{2D}$ and $\Upsilon_x$ increase only slightly. These results indicate the trend to order in the time direction, given that spatial order has already been achieved. We  notice that  $\chi_S$ has only a broad peak. This is due to the fact that, given a finite $N_{\tau}$, we see a 2D to (2+1) D crossover rather than a transition, as mentioned earlier. We have checked and found that $\chi_S$ sharpens as a function of $K_{\tau}$ or larger $N_{\tau}$.

We study the direct transition from the disordered phase to ordered phase, by choosing $K_{\tau}=0.3$ and varying $K$. The various panels in Fig. \ref{fig:a0d2o} show a compendium of results, including the  order parameter correlation functions and the relation of the spatial and the temporal correlation lengths. Both spatial and temporal ordering develop at a critical value $K_c$ between 0.45 and 0.5. The spatial and temporal correlations show similar asymptotic behaviors near the critical point,  $\propto \exp(-s/\xi_s)$ where $s=x$ or $t$. 
We find $\xi_x \sim \xi_\tau$ near the transition. This indicates, as expected for $\alpha =0$,  that the quantum transition has a dynamic critical exponent $z=1$, i.e it belongs to the classical 3d XY universality class.

\section{Complete phase diagram}
We recall the phase diagram calculated earlier in the $K-\alpha$ plane \cite{Stiansen-PRB2012, ZhuChenCMV2015} in which $K_{\tau}$ was kept fixed at a few low values. The phase diagram appears similar to Fig. (\ref{Fig-phdia-a=0}-A) with $\alpha$ replacing $K_{\tau}$. But in the $K-\alpha$ plane, the 2D quasi-ordered phase appears via a true Kosterlitz-Thouless transition as verified in I by the size dependence of the helicity modulus at the transition as well as by the correlation function of the order parameter. 

\begin{figure}[tbh]
\centering
\includegraphics[width=0.7\columnwidth]{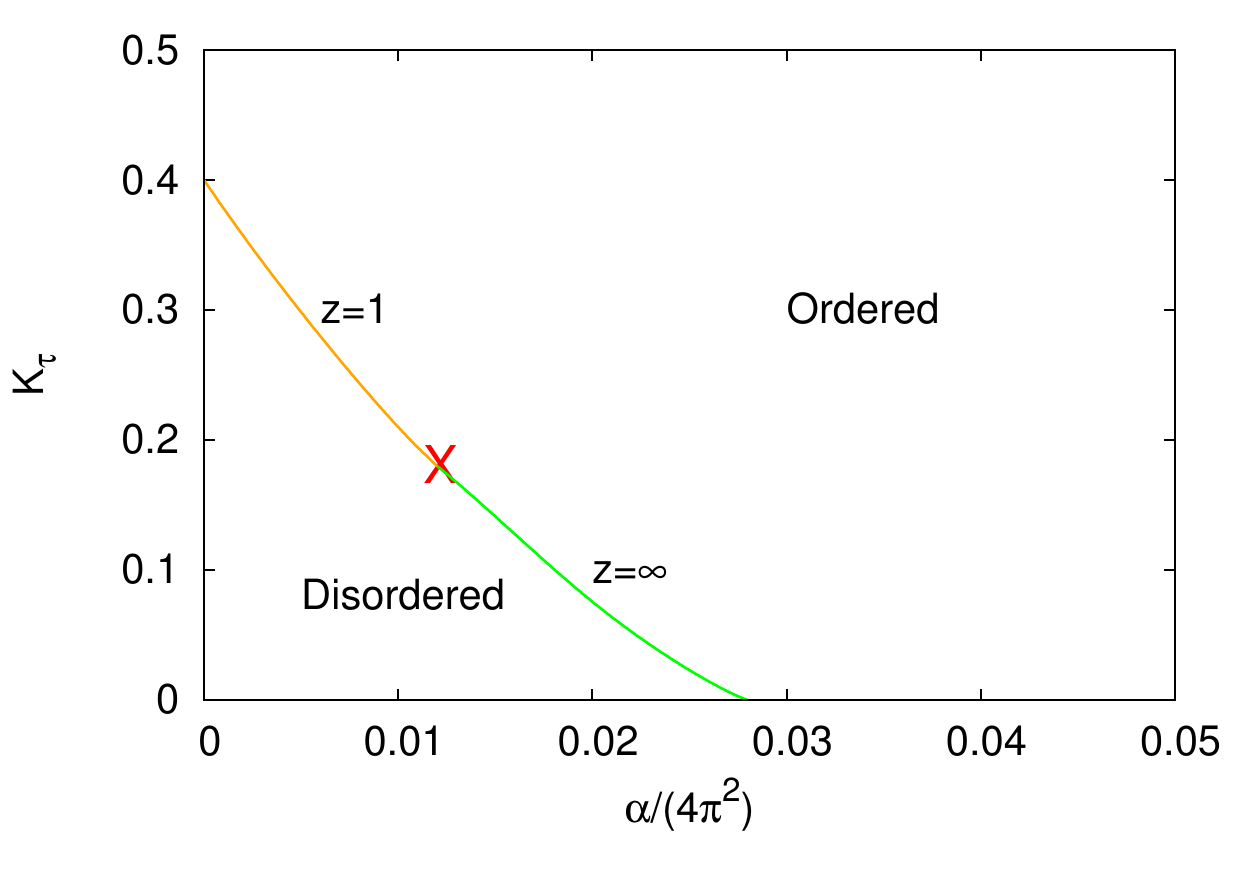}
\caption{Phase diagram in $\alpha-K_{\tau}$ plane when $K=0.4$. The red X marks the position where the transition changes from $z=1$ 3DXY university class to $z=\infty$ local critical type. Such calculations have also been done for other values of $K$ to demarcate the transition between $z=1$ and $z=\infty$ criticality shown in the phase diagram in Fig. (\ref{fig:xxx2}).}
\label{fig:K04pd}
\end{figure}

We have extended these calculations to other values of $K$ and $K_{\tau}$ and $\alpha$, which are necessary to discover the change from $z =1$ to $z \to \infty$ in the phase diagram. A compendium of results from some of the new calculations are shown in the various panels of Figs. (\ref{fig:K04Kt02}), (\ref{fig:K04Kt015}), and (\ref{fig:K04pd}). 
We show the static quantities, $M, M_{2D}, \Upsilon_x$ for various values of the parameters, as well as the  order parameter correlations, $G(x,\tau)$ for fixed time as a function of $x$ and for fixed $x$ as a function of $\tau$. From the correlation functions, the spatial correlation length $\xi_x$ and the temporal correlation length  $\xi_{\tau}$ and the relation between them are deduced. 

\begin{figure}[tbh]
\centering
\includegraphics[width=0.8\columnwidth]{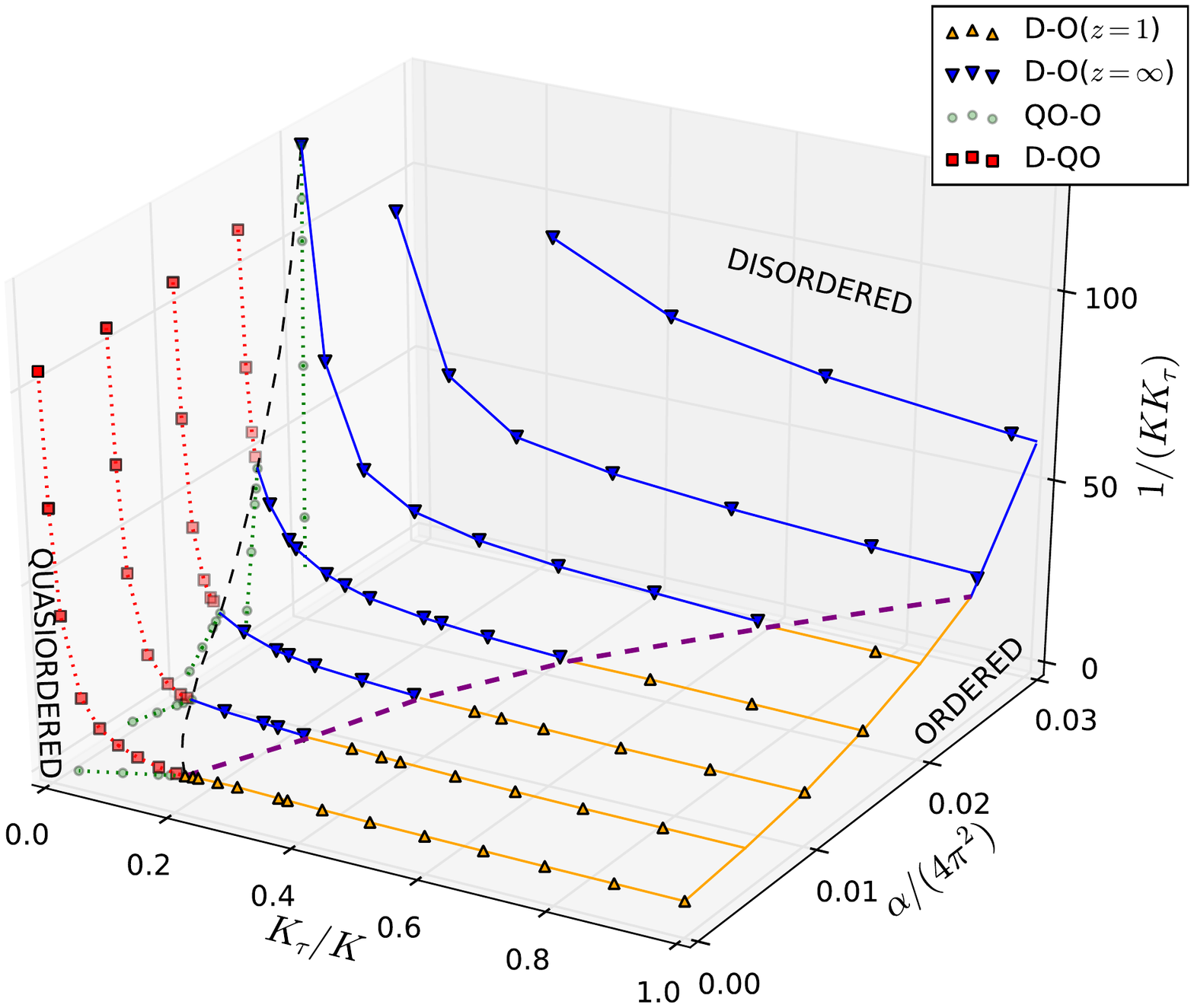}
\caption{The phase diagram in $K_{\tau}/K-\alpha-1/(K_{\tau}K)$ space. The parameter space for the Disordered, Ordered and Quasi-ordered states are specified.  Purple dots show the Disordered-Ordered transition in $z=1$ class while blue dots represent the same transition in $z \to \infty$ class. The transitions or crossovers between the Quasi-ordered to Ordered (green) and Quas-iordered to Disordered (red) are also shown. }
\label{fig:xxx2}
\end{figure}

From the results in these figures as well as calculations with other parameters, we deduce the set of transition lines (and cross-overs) between the three different phases in the $1/(KK_{\tau})-\alpha-K_{\tau}/K$ space in Fig. (\ref{fig:xxx2}).  We find that the common point of the three phases in the $1/(K_{\tau}K) - \alpha$ plane changes for a fixed value of $K = 0.4$ from  $1/KK_{\tau} \approx 12$ for $\alpha/4\pi^2 \approx 0.01$, $1/KK_{\tau} \approx 25$ for $\alpha/4\pi^2 \approx 0.018$ and for $1/KK_{\tau} \approx 250$ for $\alpha/4\pi^2 \approx 0.026$. 

There is only one transition, from the disordered to the (2+1)D ordered phase, for $K_{\tau} \gtrsim K$ and $1/KK_{\tau} \lesssim 10$.  
A plot of $\xi_x$ against $\xi_{\tau}$ is also shown, from which we find that this transition changes  from one with
$z=1$ to that consistent with $z\to \infty$ for $\alpha/4\pi^2 \gtrsim 0.01$.

The change of $z$ in the $K_{\tau}-\alpha$ plane for a fixed value of $K$ is shown in Fig. \ref{fig:K04pd}. At the small value, $K =0.4$ for which the results are shown, there is no 2D quasi-ordered phase. The transitions occur, from the disordered phase when both $K_{\tau}$ and $\alpha$ are small, to the ordered phase for larger values of these parameters. However, depending on whether the critical value of $K_{\tau}$ is smaller than or larger than about 0.18, the transition has $z=1$ or $z \to \infty$. We have evidence that the change occurs abruptly rather than smoothly. We show, in Figs.~\ref{fig:K04Kt015} and \ref{fig:K04Kt02}, the details of these two types of transitions, by two examples with $K_{\tau}$ fixed at 0.15, and 0.2 while $\alpha$ is varied. Similar results are also obtained for varying $K_{\tau}$ with fixed $\alpha$. We also conclude from these and other results that the transition from the disordered to the ordered phase remains in the $z=1$ class for $1/(KK_{\tau}) \lesssim 12$ for any value of $\alpha$. This may be significant in relation to some of the experiments in superconductor-insulator transitions, in which $z=1$ and $\nu=2/3$ has been deduced \cite{GoldmanRev2014, KapSIT2}. One should note however that exponents varying from these are also obtained in different samples. We would suggest deducing the systematics of $KK_\tau$ in experiments as samples are varied.

 \begin{figure}[tbh]
\centering
\includegraphics[width=1.0\columnwidth]{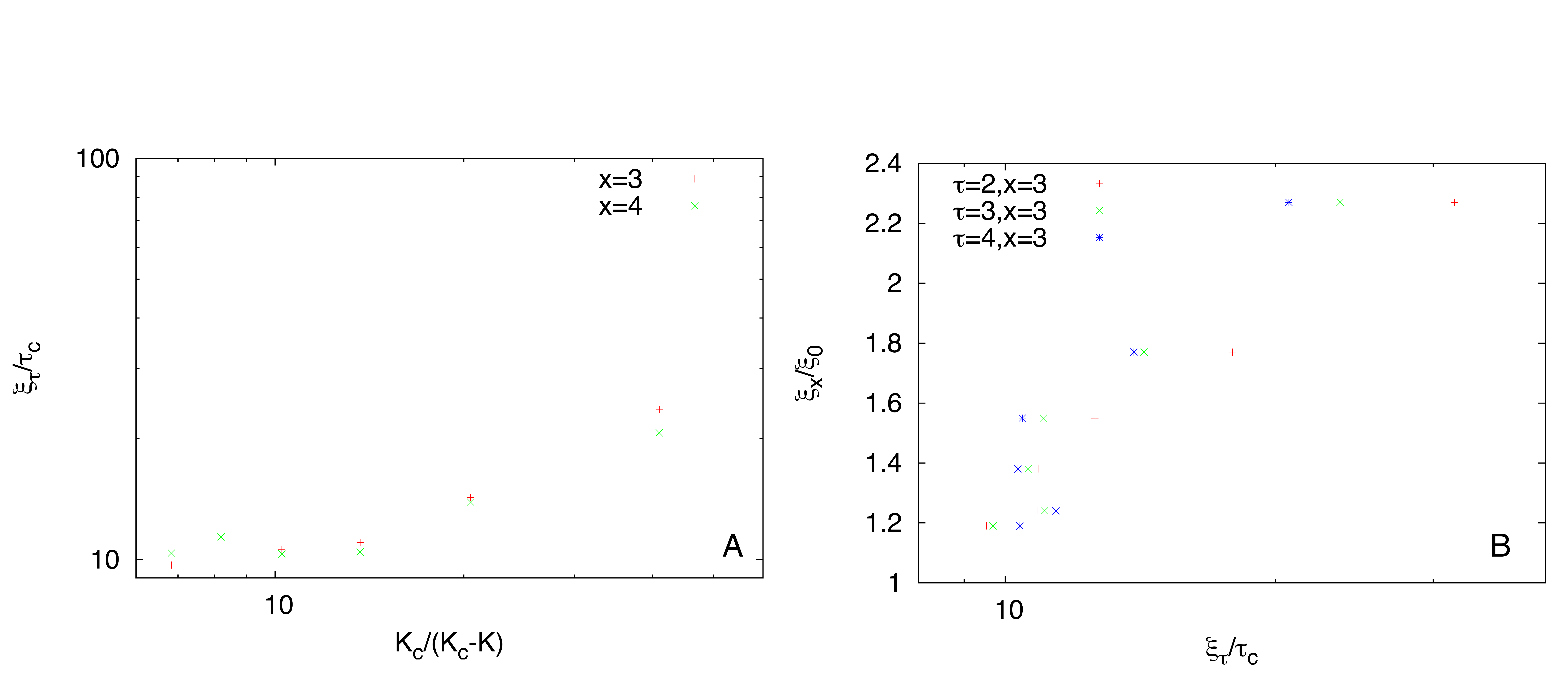}
\caption{Correlation lengths for transition driven by variation in the ratio of potential energy to kinetic energy, $KK_{\tau}$, for fixed dissipation. \\ A: The temporal correlation length $\xi_{\tau}$ as a function of $(K_c -K)$ for a fixed value of $K_{\tau}$.
Similar results are obtained as a function of $K_{\tau}$ for a fixed $K$. The critical value of the product $KK_{\tau}$ systematically decreases as the dissipation parameter $\alpha$ increases, as shown in the phase diagram of Fig. (\ref{fig:xxx2}).\\
B: The spatial correlation length $\xi_x$ increases very slowly as a function of $(K_c -K)$ compared to the temporal correlation length $\xi_{\tau}$. We draw $\xi_x$ on a linear scale and $\xi_{\tau}$ on a logarithmic scale to show that their relation appears consistent with  $\xi_x \propto \log \xi_{\tau}$}
 \label{fig:corrK}
\end{figure}

\section{Correlation functions at the Disordered to Ordered \\ transition as a function of $(KK_{\tau})$}
In I we studied the correlation function at the disordered to (2+1)D ordered transition as a function of the dissipation parameter $\alpha$ for fixed $K$ and $K_{\tau}$. We provide results in Fig. (\ref{fig:corrK}) for 
 the order parameter correlation function when the transition at large fixed $\alpha$ is driven by increasing the product $\overline{K}^2 \equiv (KK_{\tau})$. We recall that as a function of the deviation of the dissipation parameter $\alpha$ from its critical value $\alpha_c$ for fixed $K$ and $K_{\tau}$, $\xi_{\tau}$ has an essential singularity, and the spatial correlation length is consistent with varying as the logarithm of the temporal correlation length, i.e $z \to \infty$. On the other hand, we find that for fixed $\alpha$ and $K_{\tau}$, the behavior as a function of $K$ is
 \be
 \label{corrK}
 \xi_{\tau}/\tau_c = \left(\frac{K_c-K}{K_c}\right)^{-\nu_{\tau}},
 \ee
 with $\nu_{\tau} \approx 0.5$. But, as also shown in Fig. (\ref{fig:corrK}), the very slow variation of $\xi_x$ compared to $\xi_{\tau}$, consistent with $\xi_x/a \propto \log \big(\xi_{\tau}/\tau_c\big)$ 
 continues to hold. Similar exponent is found for variation of $K_{\tau}$ for fixed $\alpha$ and $K$. The calculation show that the critical $\overline{K}_c = \sqrt{KK_{\tau}}_c$ continuously decreases towards 0 as $\alpha$ becomes larger and larger. This is also exhibited in Fig. (\ref{fig:xxx2}). 
 
In experimental systems, ranging from the ferromagnetic to antiferromagnetic and loop-order quantum criticality, the variation of the parameter $(KK_{\tau})$ is more likely to drive the transition rather than the dissipation parameter $\alpha$. The superconductor-insulator transitions may be driven by either parameter. The result that a transition can occur for  smaller values of $\sqrt{KK_{\tau}}$ as $\alpha$ increases is  a very important  verifiable prediction for experiments.

\begin{figure}[tbh]
\centering
\includegraphics[width=1.0\columnwidth]{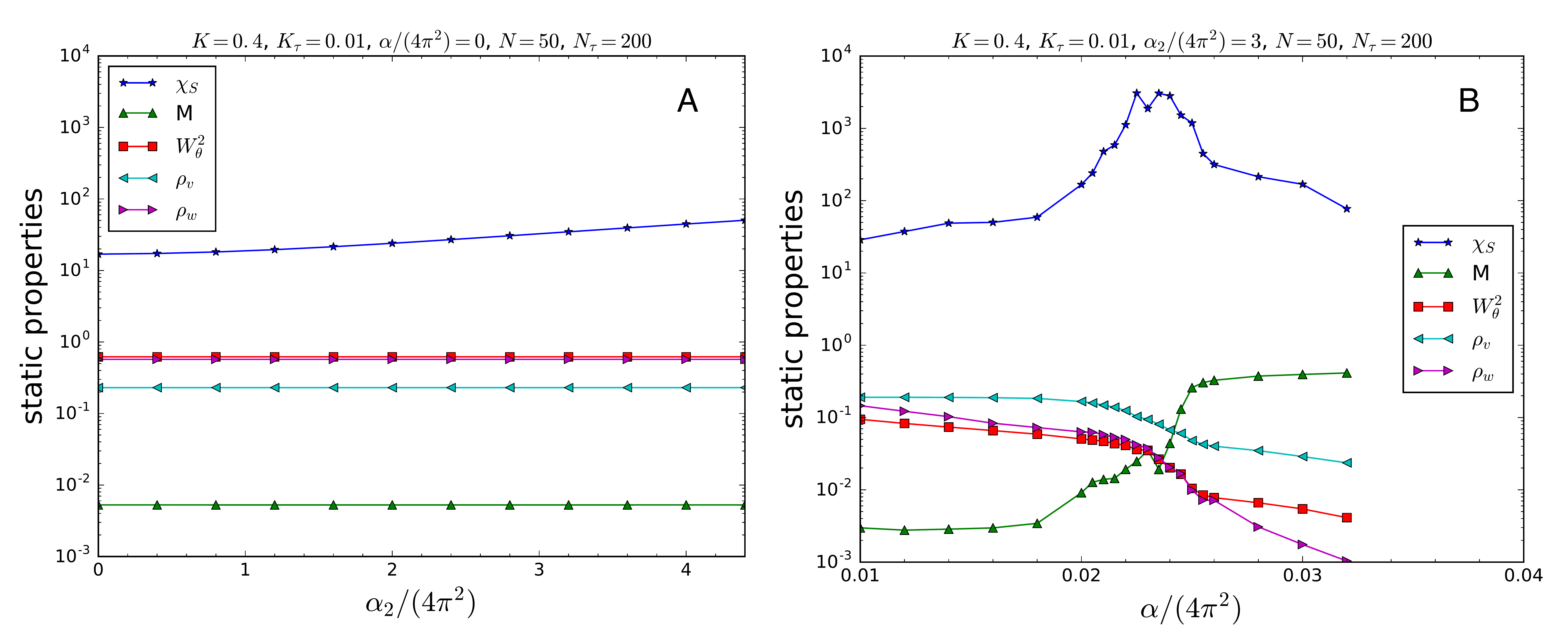}
\caption{(A): Static quantities calculated for varying values of  the parameter $\alpha_2$ for the periodic form of dissipation, with the parameter $\alpha$ for Caldeira-Legett form of dissipation kept at 0. This calculation (and a few others not shown) are in the range below the critical $KK_{\tau}$ in which there is a transition for $\alpha_2 =0$. In this range, it is shown that  there is no transition caused by increasing $\alpha_2$. (B): Static quantities calculated for a large fixed value of $\alpha_2$ as a function of  $\alpha$. Comparison with the same quantities calculated in Fig. (6) of I for $\alpha_2 =0$ shows that a finite value of the latter has no observable consequences for the nature of the transition}
 \label{fig:alphaalpha2}
\end{figure}

\section{Disorder to order transition for different forms of dissipation}

We have presented calculations above and in I with the Caldeira-Leggett form of dissipation - the third term in the action, Eq. (\ref{model}). This form of dissipation is clearly allowed even though it does not preserve the periodicity 
$\theta({\bf x}, \tau) \to \theta({\bf x}, \tau) + 2\pi$. It represents the coupling of the collective mode current, proportional to $\nabla \theta$, to fermion currents in the metallic models of interest. Another form of allowed dissipation preserves periodicity:
\be
\label{p-diss}
S_{2,diss} =  \frac{\alpha_2}{4\pi^2} \sum_{\langle{\bf x, x}'\rangle} \int d \tau  d\tau' \frac {\pi^2}{\beta^2} \frac {\cos \Big(\big(\theta_{{\bf x}, \tau} - \theta_{{\bf x}', \tau})  -(\theta_{{\bf x}, \tau'} - \theta_{{\bf x}', \tau'}\big)\Big)}{
\sin^2\left(\frac {\pi |\tau-\tau'|}{\beta}\right)} 
\ee
Calculations with, in effect, periodic dissipation alone have been presented earlier \cite{Sperstad2011}. The result is that the transition with and without such a dissipation remains in the same class, i.e. with $z=1$. We present in Fig. (\ref{fig:alphaalpha2}) calculations with both forms of dissipation present and for for kinetic and potential energy parameters fixed so that the transition studied is from the disordered state to the fully ordered state in the range in which increasing $\alpha$ leads to the transition in the $z \to \infty$ class. The two parts of the Fig. (\ref{fig:alphaalpha2}) are: (A) Varying $\alpha_2$ for fixed $\alpha = 0$. We have found that these results remain unchanged for $\alpha$ less than about the critical value  in the absence of $\alpha_2$. (B) With fixed large $\alpha_2$ varying $\alpha$ across the critical value $\alpha_c$ for $\alpha_2 =0$.  On comparison with results presented in Fig. (6) of I, the conclusion from (A) is that  there is no transition as a function of $\alpha_2$ for a very wide range of its variation for $\alpha = 0$. The conclusion from (B) is that
all the quantities calculated have precisely the same form on either side of the transition as a function of $\alpha$ as in the calculation with $\alpha_2 =0$. In fact, even the location of the transition as a function of $\alpha$ does not appear to change. 

Together with earlier results \cite{Sperstad2011}, the conclusion is that for small enough $\alpha$, the disorder to (2+1)D order transition as a function of   $KK_{\tau}$, for any reasonable value of the periodic dissipation $\alpha_2$,  retains the $z=1$ class as for $\alpha_2 =0$ . On the other hand $\alpha_2 \ne 0$ does not affect the transition for the range of $KK_{\tau}$ in which it is driven by increasing the Caldeira-Leggett dissipation $\alpha$.

{\it Acknowledgements}: 
We thank Professor Yan Chen for permission to use the computing facilities at Fudan University for some part of this work. Discussions with Vivek Aji and Asle Sudbo are gratefully acknowledged.
This work was partially supported by funds by the National Science Foundation grant DMR-1206298.

\bibliographystyle{apsrev4-1}
\bibliography{REFcopy.bib}

\end{document}